\documentclass[
    aps,      
    prx,      
    reprint,  
    noeprint, 
    nofootinbib, 
    superscriptaddress, 
]{revtex4-2}
\usepackage{physics, siunitx, mathtools, amssymb, bm}
\usepackage{xr-hyper}
\usepackage{hyperref}
\usepackage{cleveref} 

\crefname{supp:fig}{S.}{S.}

\begin{document}
\title{Observation of Critical Current Minimum in Super-Honeycomb Josephson Junction Arrays}

\author{Melissa Mikalsen}
\email{mem9982@nyu.edu}
\affiliation{Center for Quantum Information Physics, Department of Physics, New York University, New York, NY 10003, USA}

\author{Alexander-Georg Penner}
\affiliation{Dahlem Center for Complex Quantum Systems and Fachbereich Physik, Freie Universität Berlin, 14195 Berlin, Germany}

\author{Samuel D. Escribano}
\affiliation{Department of Condensed Matter Physics, Weizmann Institute of Science, Rehovot 7610001, Israel}

\author{Nadav Drechsler}
\affiliation{Department of Condensed Matter Physics, Weizmann Institute of Science, Rehovot 7610001, Israel}

\author{Arunav Bordoloi}
\affiliation{Center for Quantum Information Physics, Department of Physics, New York University, New York, NY 10003, USA}

\author{Jacob Issokson}
\affiliation{Center for Quantum Information Physics, Department of Physics, New York University, New York, NY 10003, USA}

\author{Yuval Oreg}
\affiliation{Department of Condensed Matter Physics, Weizmann Institute of Science, Rehovot 7610001, Israel}

\author{Javad Shabani}
\email{corresponding author, jshabani@nyu.edu}
\affiliation{Center for Quantum Information Physics, Department of Physics, New York University, New York, NY 10003, USA}

\date{\today}
\begin{abstract}
    Superconductor-semiconductor Josephson junction arrays are a uniquely tunable platform for studying collective quantum phenomena, particularly in the regime where localized Andreev bound states can hybridize across the lattice when the physical separation between adjacent junctions is smaller than their coherence length ($\xi_{\text{ABS}}>d_{\text{JJ}}$). Here, we investigate three distinct Al-InAs Josephson junction arrays: a square array and super-honeycomb array fabricated within this ${\xi_{\text{ABS}}>d_{\text{JJ}}}$ regime, as well as a larger-spacing super-honeycomb control device designed such that $\xi_{\text{ABS}}\lesssim\!~d_{\text{JJ}}$. Under an out-of-plane field, critical current peaks emerge at rational filling factors, reflecting stable vortex configurations in the lattices. In the super-honeycomb lattice, vortices localize to distinct non-identical plaquettes at different filling factors, as predicted by frustrated XY model simulations. A rotating in-plane field yields periodic critical current oscillations that reflect the Rashba spin-orbit coupling inherent to the InAs quantum well. Surprisingly, at $f = 1$, the closely spaced super-honeycomb array exhibits a distinct critical current minimum as the magnitude of the in-plane field increases, a signature absent in the square array and large-spacing super-honeycomb array. These results indicate that this signature is jointly influenced by the unique geometry of the super-honeycomb vortex lattice and by long-range inter-junction hybridization.
\end{abstract}
\maketitle


Josephson junction arrays (JJAs) have been extensively investigated for decades as platforms to study phenomena such as the Berezinskii-Kosterlitz-Thouless transition~\cite{resnick1981-BKT,cosmic2020probing-BKT,van1987phase,abraham1982resistive,martinoli2000two} and quantum phase transitions~\cite{newrock2000two,fazio2001quantum,bottcher2018anomolous}. Junction arrays have been classically modeled with the XY model~\cite{teitel1983phase-xy,ashrafuzzaman2003BKT-XYmodel}. These investigations have utilized conventional superconductor-insulator-superconductor (SIS) arrays~\cite{van1992field,trias1995self,fazio2001quantum,martinoli2000two,newrock2000two} and modern superconductor-semiconductor (S-Sm) platforms~\cite{bottcher2018anomolous,bottcher2023dynamical,bottcher2024-BKT,sasmal2025voltage-tune-channel-gate,reinhardt2025-diode-array,baumgartner2021-1DJJA-B_par}. Applying an out-of-plane magnetic field $B_{\perp}$ induces a phase accumulation around each plaquette, quantified by the frustration parameter $f=B_{\perp}A_{\text{uc}}/\Phi_0$, where $A_{\text{uc}}$ is the unit cell area and $\Phi_0=h/2e$ is the flux quantum~\cite{teitel-jayprakash1983,tinkham1983periodic-vortex}. This frustration stabilizes periodic vortex configurations whose commensurability depends on the underlying lattice geometry~\cite{halsey1985-square-lattice-theory,lin2002-varyingJJA-patterns,park2001-kagome-JJA-theory}. Beyond square structures, more complex array lattices have also been explored, such as multi-terminal networks~\cite{teller2025frustrated-frustration} and dice lattices~\cite{bondar2025-dice}.

While the geometric constraints governing commensurate vortex lattices are largely independent of the individual junction platform, substituting insulating barriers with semiconducting weak links changes the mechanism of supercurrent transport. In conventional SIS networks, Cooper pairs tunnel through the insulating barrier where the superconducting wavefunctions decay and overlap \cite{tinkham2004introduction}. In contrast, supercurrent in S-Sm junctions is carried by Andreev bound states (ABS) arising from coherent reflections of electrons and holes at the S-Sm interface~\cite{andreev1965thermal-original-andreev-reflection,blonder1982-ABS}. These ABS are localized in the semiconducting weak link and decay into the superconducting leads. Here, the relevant length scale is the Andreev bound state coherence length, $\xi_{\rm ABS}=\xi/(\sqrt{T_N}|\sin{\varphi/2}|)$, where $T_N$ is the transparency and $\xi$ is the coherence length of the superconducting proximity effect~\cite{prada2020andreev-abs-coherence-length}. When the physical spacing between adjacent junctions ($d_{\text{JJ}}$) is small enough that neighboring ABS overlap, $\xi_{\text{ABS}}\gg d_{\text{JJ}}$, they can hybridize to form Andreev molecules~\cite{pillet2019nonlocal-Andreev-molecules,pillet2020scattering-andreev-molecules,Kornich-overlapping-ABS-andreev-molecules,haxell2023demonstration-andreev-molecules-exp,matsuo2023phase-andreev-molecule-exp}. Coupling S-Sm junctions in an array that satisfies this length scale has been described as an Andreev crystal~\cite{dahl2025andreev-crystal}, where ABS hybridize across the entire array.

Implementing this S-Sm platform also enables an in-plane magnetic field to act as a tuning mechanism, leveraging the semiconductor's intrinsic spin-orbit coupling. For near-surface InAs quantum wells \cite{shabani2016AlInAs},  the spin-orbit length scale ($l_{\text{SO}}$) is estimated to be $300 \text{ nm}$ \cite{farzaneh2024observing}. An in-plane field has been shown to modify a S-Sm array's quantum phase transitions~\cite{bottcher2024-BKT}, as well as skew the vortex pinning potential landscape~\cite{reinhardt2025-diode-array}. The combined effects of structural design, out-of-plane frustration, and in-plane magnetic field are therefore central to understanding S-Sm Josephson junction arrays in the regime where neighboring junctions are not independent. This parameter space is of particular interest given a recent proposal indicating that such arrays can be carefully tuned into a two-dimensional chiral topological superconducting phase \cite{lesser2024josephson}.


Here we report measurements of Josephson junction arrays fabricated from an Al-InAs heterostructure \cite{shabani2016AlInAs}, to experimentally investigate the combination of structural design and field tunability in S-Sm platforms. We investigate square and super-honeycomb arrays with length scales such that $\xi_{\rm ABS}$ is comparable to or larger than $d_{\text{JJ}}$, and compare them with a larger-spacing super-honeycomb control device. Magnetotransport measurements of the critical current $I_c$ versus out-of-plane magnetic field reveal geometry-dependent vortex configurations. We show that variations of $I_c$ under an angled in-plane magnetic field reflect the lattice geometry and the semiconductor's Rashba-dominated spin-orbit interaction. In addition, $I_c$ versus in-plane field at $f=1$ reveals a surprising minimum for the closely spaced super-honeycomb array. This feature is absent in the square array and in the larger-spacing super-honeycomb control device. 

Numerical calculations, based on a model that includes energy minimization of the XY model in the presence of the $\varphi_0$ effect due to spin-orbit coupling with in-plane field, reproduce stable vortex configurations for the super-honeycomb lattice. Application of an external current induces disorder in the lattice when it is close to the critical current.  Applying an in-plane field causes a $\varphi_0$ effect that reduces the critical current, which aligns with the experimental dependence of the critical current on the magnitude of the applied in-plane field.

The inclusion of the $\varphi_0$ effect does not reproduce the observed minimum critical current for $f=1$. This discrepancy may be due to the necessity of incorporating the hybridization between the junctions when $\xi_{\rm ABC} > d_{\text{JJ}}$.


\begin{figure*}[tbp]
    \centering
    \includegraphics{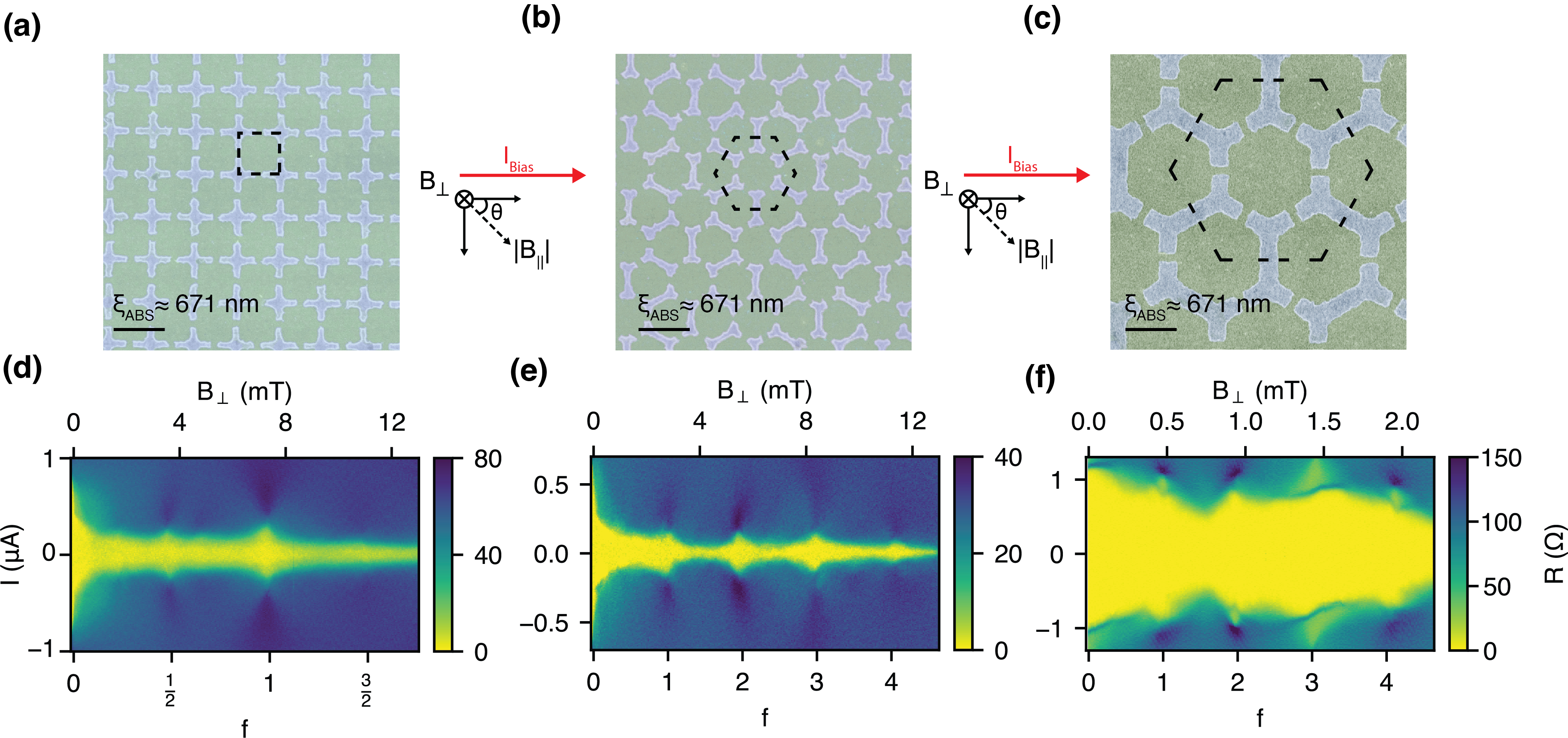}
    \caption{
        \textbf{Josephson junction arrays in perpendicular magnetic field.} Scanning electron micrograph of a few plaquettes of a representative device for the square \textbf{(a)} and the super-honeycomb array with $\xi_{\text{ABS}} > d_{\text{JJ}}$ \textbf{(b)} and $\xi_{\text{ABS}} \lesssim  d_{\text{JJ}}$ \textbf{(c)}. The superconductor (blue) and semiconductor (green) are falsely colored, the unit cell is outlined (dashed black line), and the scale bar is 671 nm. We plot the resistance as a function of current bias (I ($\mu A$) ) and frustration ($f = B_{\perp} A_{\text{UC}}/ \Phi_0$) for the square \textbf{(d)} and the super-honeycomb array with $\xi_{\text{ABS}} > d_{\text{JJ}}$ \textbf{(e)} and $\xi_{\text{ABS}} \lesssim d_{\text{JJ}}$ \textbf{(f)}. 
    }
    \label{fig:unit cell}
\end{figure*}

We investigate planar S-Sm (Al-InAs) Josephson junction arrays of two distinct geometries: square (Fig.\ref{fig:unit cell}~(a)) and super-honeycomb (Fig.\ref{fig:unit cell}~(b, c)). The near-surface InAs quantum well confines a two-dimensional electron gas and is superconducting through proximity to the epitaxial Al layer\cite{shabani2016AlInAs,wickramasinghe2018transport-AlInAs}. The square array (Fig.\ref{fig:unit cell}~(a)) and small super-honeycomb array (Fig.\ref{fig:unit cell}~(b)) are fabricated such that the distance between two junctions is less than the estimated Andreev bound state coherence length, thus in an Andreev crystal regime. The large super-honeycomb control device (Fig.\ref{fig:unit cell}~(c)) is designed to be outside of this regime. Here, we estimate the Andreev bound state coherence length to be $\xi_{\text{ABS}} = \sqrt{\xi_0 l}\approx 671$nm, where $\xi_0 = \hbar v_F / \pi \Delta\approx 1.66\mu$m. Here the Josephson array gap estimated from the critical temperature is $\Delta = 1.764k_BT_c\approx 160\mu$eV and the Fermi velocity is $v_F = \hbar \sqrt{2\pi n }/m^* \approx 1.27 \times 10^6$m/s. We have obtained the mean free path $l = 271$nm and density $n=1.1\times 10^{12}{\rm cm}^{-2}$ from magnetotransport data,  the critical temperature of the array, $T_c = 1.1$K, measured during the device cooldown, and  the effective band mass of InAs is taken to be $m^* = 0.024m_e$ \cite{adachi2009-InAs-eff-band-mass, schiela2025geometric,mayer2019abs-coherence-length}. 

The square array, shown in Fig.\ref{fig:unit cell}~(a), consists of a square lattice with unit cell area ($A_
{\rm uc}$) estimated from SEM images to be approximately $0.279 \mu{\rm m}^2$, and the length of the superconducting island between two junctions is approximately $d_{\text{JJ}} =470$nm, thus $\xi_{\text{ABS}}> d_{\text{JJ}}$. 
Stable vortex configurations, with elevated critical current, emerge at rational values of the frustration parameter, $f = B_\perp A_{\rm uc}/\Phi_0$, which represents the fraction of a magnetic flux quantum threading a single unit cell~\cite{teitel-jayprakash1983}. Fig.\ref{fig:unit cell} (d) shows peaks in the critical current for stable vortex configurations for the square array, most prominently for $f = 1$, where there is one vortex at each square, and $f = 1/2$, when a stable checkerboard configuration of vortices is stabilized.  A less prominent feature also appears near $f = 2/3$, but we do not observe a similar feature around $f = 1/3$.  The reason for this asymmetry may come from the response of the bulk superconductor to the external magnetic field, which is higher at $2/3$ than at $1/3$, or the peak for $f = 1/3$ may be obscured by the broad peak at $f = 0$.

The peaks in the critical current at these rational values of $f$ align with the estimated area of the unit cell for the square lattice. Additionally, the fractional values of $f$ at which these features occur have previously been predicted \cite{lin2002-varyingJJA-patterns, halsey1985-square-lattice-theory, teitel-jayprakash1983} and observed \cite{bottcher2024-BKT, lankhorst2018annealed} in arrays of a similar square geometry.

\begin{figure*}[tpb]
    \includegraphics{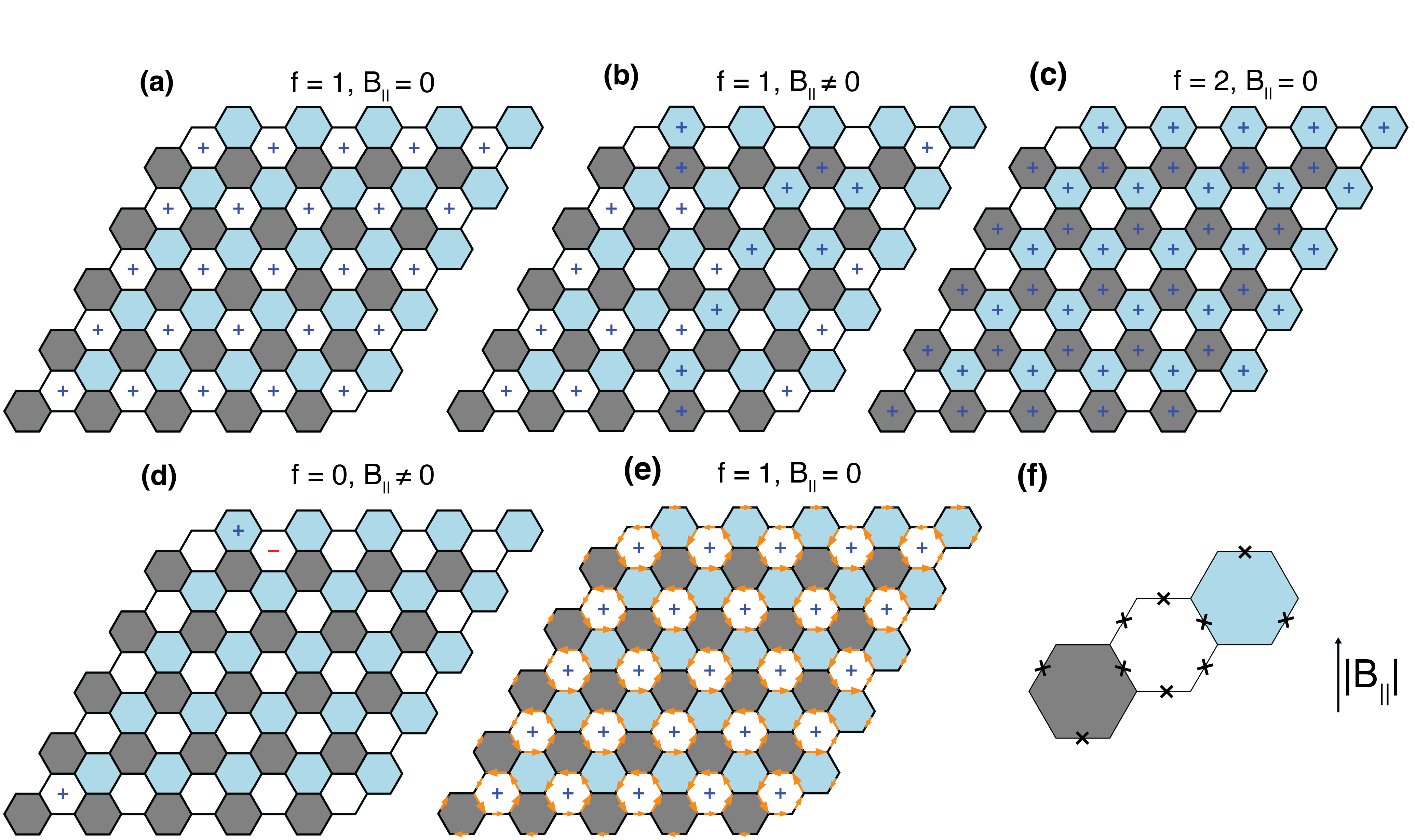}
    \caption{
        \textbf{Vortex configurations in Josephson junction arrays.} Calculated vortex configurations, for a super-honeycomb  Josephson junction array for $f = 1$ with $B_{||} = 0$ \textbf{(a)} and $B_{||} = 0.6$ \textbf{(b)}, f =2 with $B_{||} = 0$ \textbf{(c)}, and $f = 0$ wih $B_{||} = 0.6$ \textbf{(d)}. Calculation of the current distribution in the array for $f = 0$ and $B_{||} = 0$, represented by orange arrows, magnitude of the arrows represent the relative magnitude of the current \textbf{(e)}. White hexagons represent plaquettes with 6 junctions, blue and gray hexagons represent plaquettes with 3 junctions, an example unit cell is shown in \textbf{(f)}. Blue plus symbols are vortices and red minus symbols are anti-vortices. The in-plane magnetic field is pointing perpendicular to the supercurrent through 1/3 of the junctions in the array.  
    }
    \label{fig:theory-hex-vortex}
\end{figure*}

The super-honeycomb array, shown in Fig. \ref{fig:unit cell} (b, c), is a structure with a unit cell consisting of the area of three hexagonal plaquettes. Unlike in the square array, the plaquettes in the super-honeycomb array are non-identical, one third of the plaquettes host 6 Josephson junctions, while two-thirds host 3 junctions. Here, the hexagonal shape of the plaquettes maps to a honeycomb structure, while the network created by the junctions is homeomorphic to a kagome lattice (see S.\ref{supp:fig:kagome} for a visual representation of this translation). For the case of $f = 1$, vortices are predicted to be localized to plaquettes with 6 junctions because the circulating current is lower \cite{rzchowski1990vortex-pinning-potential}, while at $f = 2$, the vortices shift to being localized in the 3 junctions hexagons, see Fig. \ref{fig:theory-hex-vortex} (a,c). 
Thus, symmetry of the vortex lattice in super-honeycomb is distinct from that of the square lattice. In the square array at $f = 1/2$, uniformly adding $\Phi_0$  to every plaquette shifts the checkerboard vortex lattice to an energetically equivalent configuration. However,  for the super-honeycomb array at $f =1$,  uniformly adding $\Phi_0$  to every plaquette forces vortices from the 6 JJ plaquettes to the 3 JJ plaquettes, breaking the symmetry of the lattice. 

In Fig.\ref{fig:unit cell} (e, f), the peaks in the critical current are consistent with integer values of frustration, corresponding to the area of the unit cell outlined in Fig.\ref{fig:unit cell} (b, c), estimated to be $A_{\rm{uc}} = 0.738\mu{\rm m}^2$ for the small super-honeycomb array and $A_{\rm{uc}} = 4.39\mu{\rm{m}}^2$ for the large super-honeycomb array.  For the small super-honeycomb array in the Andreev crystal regime, $d_{\text{JJ}}\approx 235$nm  for the 6 JJ plaquette and $d_{\text{JJ}}\approx 550$ nm for the 3 JJ plaquette. However, for the large super-honeycomb array, $d_{\text{JJ}}\approx 670$nm for the 6 JJ plaquette and $d_{\text{JJ}}\approx 1.3\mu$m for the 3 JJ plaquette, therefore $\xi_{\text{ABS}}\lesssim d_{\text{JJ}}$. All measurements of the arrays were taken in a 4-point geometry from corner contacts on each array (see S. \ref{supp:fig:full-square-array} and S. \ref{supp:fig:full-hex-array}). 

In Fig.\ref{fig:unit cell} (d, e) we observe a suppression between the $0$-flux and $1$-flux fields by roughly a factor of $1/2\pi$, expected from a Fraunhofer pattern. Here, this pattern presumably originates from trajectories passing through the metallic semiconducting regions at the center of the plaquettes. Additionally, for all arrays presented, when the critical current is large, the normal-state resistance is also large. One possible explanation is correlations among vortices: if vortices move collectively, their motion may lead to enhanced dissipation and, therefore, larger resistance.

\begin{figure}
    \centering
    \includegraphics{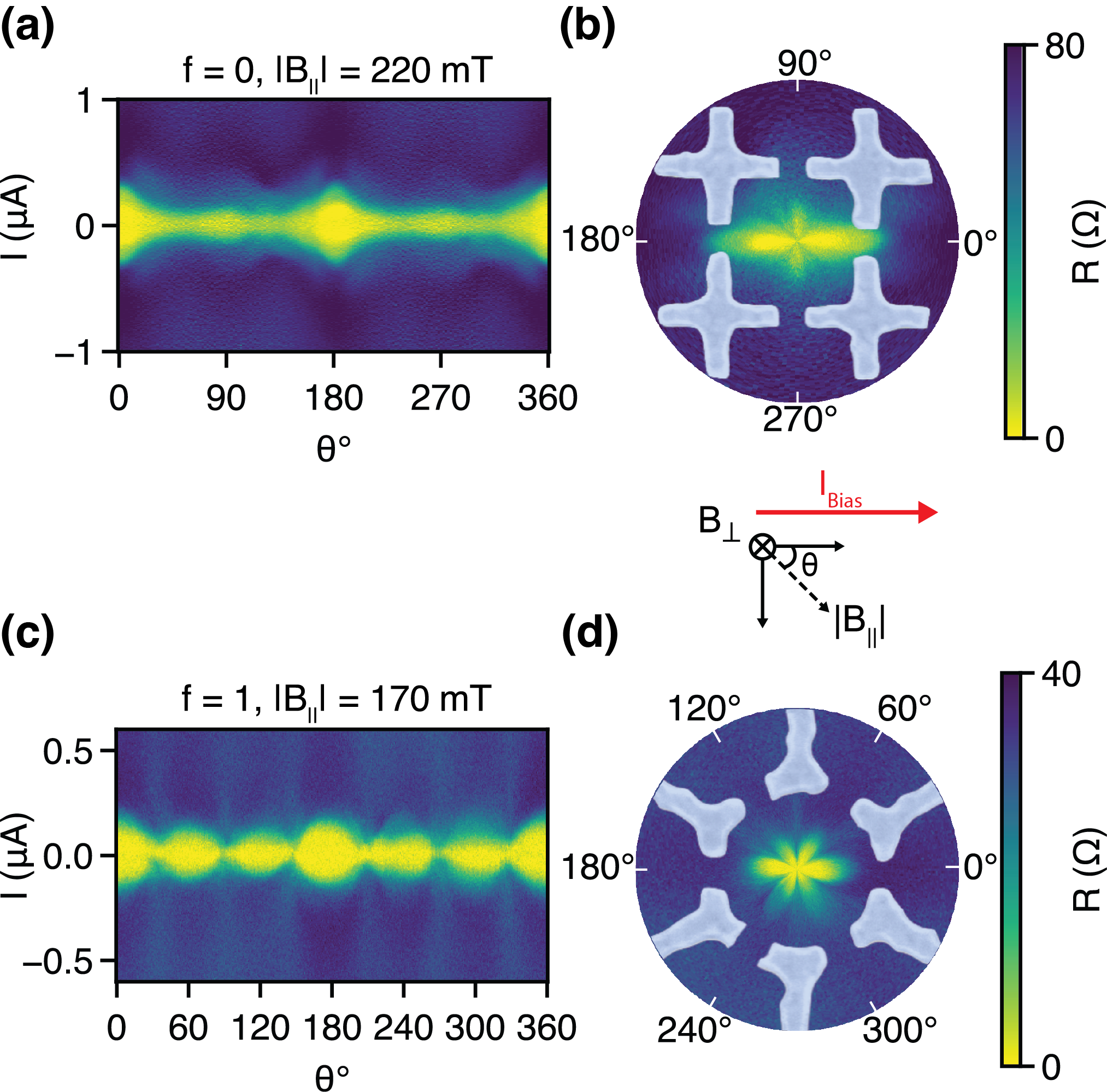}
    \caption{
        \textbf{In-plane magnetic field angle dependence of critical current.} Resistance, R ($\Omega$) as a function of current bias, I ($\mu$A), and angle of in-plane magnetic field, $\theta$. Measurements for the square array were taken at $f = 0$, corresponding to $B_\perp$ = 0 mT, and a magnitude of in-plane field of $B_{||}$ = 220 mT \textbf{(a,b)}. For the super-honeycomb array, measurements were taken at $f = 1$, corresponding to $B_{\perp}$ = 2.802 mT, and a magnitude of in-plane field of $B_{||}$ = 170 mT  \textbf{(c,d)}.  The data from both array geometries is plotted on a polar plot with the corresponding representative unit cell SEM image overlaid. Only the data from the positive current bias is plotted on the polar plot for both the square array\textbf{(b)} and the super-honeycomb array \textbf{(d)}. Bias current in the array is along the 0-180 deg.  
        \label{fig:angle}
    }
\end{figure}

We now study the dependence of the critical current on the angle of the in-plane magnetic field. Fig.~\ref{fig:angle} shows the resistance as a function of current bias and angle of in-plane magnetic field. The critical current reaches a maximum periodically when the in-plane magnetic field is perpendicular to the supercurrent direction in any junction within the array.
This behavior has been attributed to Rashba-dominated spin-orbit interaction, with estimated length scale $l_{\text{SO}}\approx 300\text{nm}$ \cite{farzaneh2024observing}, larger than the average length of a junction in the array ($l_{\text{JJ}} \approx 80 \text{nm}$). This interaction couples most strongly to magnetic fields perpendicular to the quasiparticle momentum \cite{fu2008superconducting,pientka2017topological}. This has been observed in single junctions as a maximum in $I_\text{c}$ when the in-plane magnetic field is perpendicular to the supercurrent \cite{schiela2025gate-3-fold-rotation-symmetry}. For the square array, this dependence is most prominent for $f = 0$ at high in-plane field, see Fig.\ref{fig:angle} (a, b). See S.\ref{supp:fig:square-angle} for critical current as a function of angle for $f = 1$ and $f = 1/2$.

In Fig.\ref{fig:angle} (a, b) the magnitude of the critical current is largest at $0^\circ$ and $180^\circ$ and smaller at $90^\circ$ and $270^\circ$, while the current flows along the $0-180$ line. This asymmetry could arise from the way the bias current is injected into the array, with the source at the bottom-left contact and the drain at the bottom-right contact; see S.\ref{supp:fig:full-square-array}. When the magnetic field is pointing along $0^\circ - 180^\circ$, the $I_\text{c}$ and coupling between Al islands along $90^\circ - 270^\circ$ are enhanced by the presence of spin-orbit coupling in the diffusive limit \cite{schiela2025gate-3-fold-rotation-symmetry}. 
This effectively creates large JJs in series with a contact width the size of the width of the array ($\approx$ 5 $\mu m$), and thus a larger $I_\text{c}$. However, when the magnetic field is pointed along $90^\circ - 270^\circ$, the coupling between the Al islands along $90^\circ - 270^\circ$ is suppressed. This cuts off the flow of current in the upper rows, effectively leading to JJs in series with a smaller contact width of about $80$ nm, and thus a smaller $I_\text{c}$. 

For the super-honeycomb array, the oscillations in the critical current also align with the condition where the magnetic field is perpendicular to the supercurrent through a junction in the unit cell. We again attribute this to the Rashba-dominated spin-orbit interaction in the semiconductor. This feature is most prominent for $f = 1$ at high in-plane field magnitudes, see Fig.\ref{fig:angle}~(c,d). For further investigation of the dependence of the critical current on the angle of the in-plane magnetic field, see S.\ref{supp:fig:f-0-angle} for $f = 0$ at various magnitudes of in-plane field, S.\ref{supp:fig:f-1-angle} for $f =1$ at smaller in-plane magnitudes, S.\ref{supp:fig:f-2-angle} (b) for $f = 2$ at high in-plane field and S.\ref{supp:fig:f-3-angle} (b) for $f = 3$ at high in-plane field.


\begin{figure*}[tbp]
    \centering
    \includegraphics{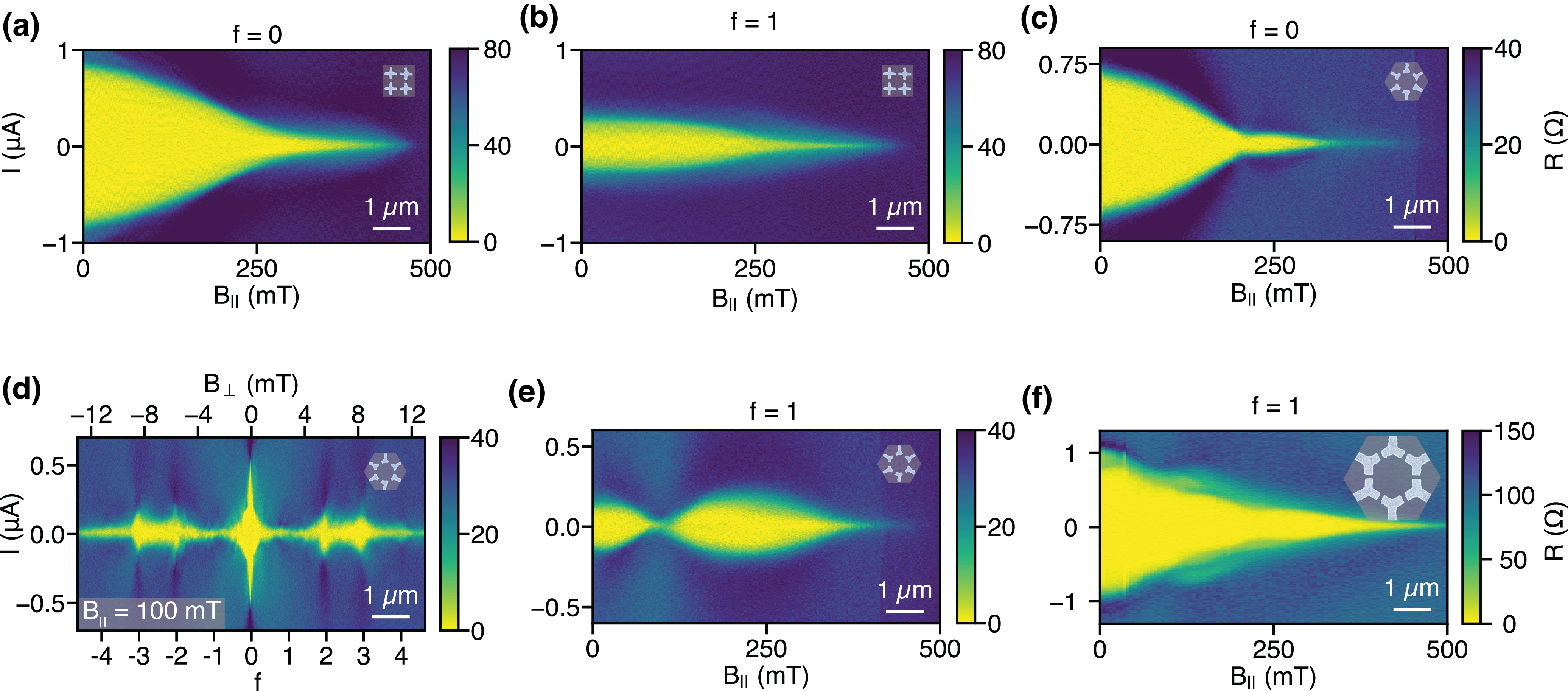}
    \caption{
    \textbf{Josephson junction arrays at finite frustration with in-plane magnetic field.} Resistance as a function of current bias (I ($\mu A$)) and in-plane magnetic field pointing along $\theta$ = 0$^\circ$ ($B_{||}$ (mT)) for the square array at constant $f = 0$ \textbf{(a)} and $f = 1$ \textbf{(b)} and the $\xi_{\text{ABS}} > d_{\text{JJ}}$ super-honeycomb array at $f = 0$ \textbf{(c)} and $f = 1$ \textbf{(e)}. Resistance as a function of current bias (I $(\mu A)$) and frustration ($f=B_{\perp} A_{\text{UC}}/ \Phi_0$) for the super-honeycomb array at constant in-plane magnetic field, $|B_{||}({\theta=0^{\circ})}|$ = 100 mT \textbf{(d)}. Resistance as a function of current bias (I ($\mu A$)) and in-plane magnetic field pointing along $\theta$ = 0$^\circ$ ($B_{||}$ (mT))  the $\xi_{\text{ABS}} \lesssim d_{\text{JJ}}$ super-honeycomb array at $f = 1$ \textbf{(f)}.
    }
    \label{fig:in-plane}
\end{figure*}

We investigate the effect of in-plane magnetic field on the arrays at zero and finite frustration. Fig.\ref{fig:in-plane}~(a,b) show the resistance as a function of current bias and in-plane magnetic field pointing along $\theta=0^\circ$ for the square array. This direction is perpendicular to the supercurrent in half of the junctions. For constant frustration, $f = 0$ (Fig.\ref{fig:in-plane}(a)) and $f = 1$ (Fig.\ref{fig:in-plane} (b)), the supercurrent decreases until it fully depletes around $475$ mT. See S.\ref{supp:fig:square-angle}
(a) for in-plane field measurements at $f = 1/2$. 

For the small super-honeycomb array at $f = 0$, shown in Fig.\ref{fig:in-plane} (c), the dependence of the critical current on the magnitude of in-plane field pointing along $\theta =0^\circ$ is qualitatively similar to that of the square array. However, at $f = 1$ in Fig.\ref{fig:in-plane} (e), there is a minimum in the critical current of the array around $100$ mT, then $I_{\text{c}}$ increases as $B_{\parallel}$ increases, before fully depleting at higher fields.  Fig.\ref{fig:in-plane} (d) shows the resistance as a function of current bias and frustration for the super-honeycomb array at a constant in-plane magnetic field of $100$ mT, the field value at which the minimum in the supercurrent occurs in Fig.\ref{fig:in-plane} (e). This shows that the peak in the critical current at $f = 1$ has been reduced, while the peaks at $f = 0, 2,$ and $3$ are preserved. Here, the in-plane field is pointing along $\theta = 0^\circ$, similar features are observed when the in-plane field is pointing along perpendicular to the supercurrent through the other sets of JJs in the unit cell at $\theta = 60^\circ$ and $\theta = 120^\circ$, see S.\ref{supp:fig:f-1-60and120deg}. 

The minimum observed in Fig.\ref{fig:in-plane} (e) could be due either to a reduction of the Josephson energy $E_J$ in an in-plane field, or possibly to a $\varphi_0$ shift arising in the presence of next-nearest-neighbor interactions. To investigate the role of length scales of the device and the regime where $\xi_{\text{ABS}}\lesssim d_{\text{JJ}}$, we investigate a super-honeycomb array with the same number of plaquettes, but with a unit cell area $A_{\text{uc}} = 4.39\mu {\rm m}^2$, as shown in Fig.\ref{fig:unit cell} (c). Unlike in the Andreev crystal regime at $f = 1$, no full suppression and re-emergence of the supercurrent was observed in Fig. \ref{fig:in-plane} (f). While this minimum is prominent in the small super-honeycomb array, it is suppressed in the large array and absent in the square lattice. This contrast indicates that the feature is influenced by both the device's small length scales and the spatial symmetry breaking of the vortex lattice.

We note that the minimum in the critical current as shown in Fig.\ref{fig:in-plane} (e) likely cannot be explained by a Fraunhofer interference pattern. The length of a single junction in the array is approximately $80$ nm, and the quantum well thickness is effectively 12 nm. This gives an estimate of the first node in a Fraunhofer pattern to be on the order of a few tesla, more than an order of magnitude larger than the minimum observed at about 100 mT. Moreover, this Fraunhofer node would be observed independent of geometry and filling factor, which not the case for the arrays investigated. 

To understand the geometry-dependent vortex configurations and the in-plane-field-induced disorder, we model the Josephson junction array in the experiment through a frustrated 2D XY model as follows, 
\begin{equation}
    H = -E_J\sum_{\langle ij\rangle} \cos(\varphi_i -\varphi_j -2\pi aA_{ij}/\phi_0-\varphi_0^{ij}),
    \label{eq: xy model}
\end{equation}
with $E_J$ denoting the Josephson energy, $a$ the lattice constant of the array, $\Phi_0 = h/2e$ the superconducting flux quantum, and $\varphi_i$ the order-parameter phase of the $i$th superconducting island. The plaquettes of the array are threaded by a uniform
magnetic flux $\Phi$, as captured by the vector potentials $A_{ij}$ associated with the bonds between nearest-neighbor sites $i$ and $j$ of the array. Summing $A_{ij}$ around each of the plaquettes yields the flux applied to the plaquette. We quantify these fluxes in terms of the associated frustrations. The background field introduces a uniform frustration $f=\Phi/\Phi_0$. 

We search for ground-state configurations of the array as described by Eq.\ \eqref{eq: xy model}. The dynamics is simulated using the RCSJ model following the algorithm described in Ref.~\cite{lankhorst2018annealed}. The Kirchhoff rules at each lattice site result in $6N^2$ coupled linear differential equations, with $N \times N$ being the number of unit cells. Nonzero temperature enters the dynamics via Langevin currents (Johnson-Nyquist noise). The algorithm evolves the state of the system towards a minimum-energy configuration by gradually decreasing the temperature. 

The term $\varphi_0^{ij}$ represents a phase shift induced by the in-plane magnetic field, defined for each bond as \cite{buzdin2008direct}:
\begin{equation}
    \varphi_0^{ij} \propto   \frac{\alpha}{v_F^2 \hbar} E_Z L_{ij} \sin(\theta_{ij})
\end{equation}
where $E_z$ is the Zeeman energy due to the in-plane field strength, $\alpha$ is the Rashba spin-orbit coupling strength, $v_F$ is the fermi velocity,  $\theta_{ij}$ is the relative angle between the field and the junction, and structural disorder is introduced by treating the individual junction lengths $L_{ij}$ as random variables drawn from a normal distribution. This geometric randomness directly modulates the $\varphi_0^{ij}$ shift at each bond, converting a uniform in-plane field into a spatially non-uniform phase configuration that destabilizes the ordered vortex lattice.

In Fig.\ref{fig:theory-hex-vortex} (a), the vortices, represented by the blue plus symbols, are localized to the 6 junction hexagons at 0 in-plane magnetic field. While at $f = 2$, the vortices are localized to the $3$ junction hexagons, as shown in Fig. \ref{fig:theory-hex-vortex} (c). These stable vortex configurations manifest in the peaks in the critical current as seen in Fig. \ref{fig:unit cell} (e, f). 


As shown in Fig.~\ref{fig:theory-hex-vortex} (d), for $f = 0$, when an in-plane magnetic field is applied, some vortices (blue plus symbols) and anti-vortices (red minus symbols) are induced in the array. While at $f = 1$ (Fig. \ref{fig:theory-hex-vortex} (b)) at the same finite in-plane field, some vortices shift from being localized to the 6 JJ hexagons (white) to the 3 JJ hexagons (blue or gray). The number of vortices shifted in the $f = 1$ case is larger than the induced number of vortices / anti-vortices in the $f = 0$ case. Indicating that the in-plane field-induced disorder is larger for $f =1$ than for $f =0$.

The numerical simulations indicate that the in-plane magnetic field increases disorder in the vortex lattice configuration. Similarly, the data presented for the small super-honeycomb array at $f = 0$ (Fig.~\ref{fig:in-plane} (c)), and the large super-honeycomb array at $f  = 1$ (Fig.\ref{fig:in-plane} (f)) shows a decrease in $I_c$ with increasing in-plane magnetic field. However, the XY model presented here is insufficient to explain the minimum in the critical current at $f = 1$, as shown in Fig.\ref{fig:in-plane} (e). This discrepancy is presumably because the XY model presented here does not include the effect of next-nearest neighbor coupling, relevant in the limit where $\xi_{\text{ABS}}>d_{\text{JJ}}$. Therefore, the origin of this signature may be related to the length scale limit where there is coupling between adjacent plaquettes. 




In single 1D nanowires and planar JJs based on S-Sm materials, an in-plane magnetic field is predicted to drive a topological phase transition by inducing a band inversion in the semiconductor that closes and reopens the bulk superconducting gap\cite{oreg2010majorana,lutchyn2010majorana-super-semi, pientka2017mzm-planar-jj,hell2017mzm-planar-jj}. In this case, a minimum in the critical current is a predicted signature of a topological phase transition \cite{pientka2017mzm-planar-jj}, and has been observed in S-Sm single JJ devices\cite{ren2019topological-phase-bias-theory,fornieri2019-topological-exp-min-re-entrant,dartiailh2021-min-Ic}. However, this phase transition requires large magnetic field. 

To overcome the requirement for a strong Zeeman field, recent theoretical works proposed that networks of S-Sm junctions can realize a collective two-dimensional chiral topological superconducting phase through structural engineering and magnetic flux tuning~\cite{lesser2024josephson}. Such two-dimensional S-Sm JJ arrays introduce new symmetry-breaking mechanisms beyond those available in single junctions, including control of superconducting island phases, magnetic flux threading~\cite{lesser2024josephson}. The geometry and length scale of our small super-honeycomb device were inspired by these proposals, placing it in a regime where coherent coupling between neighboring junctions becomes important. While additional numerical studies we have performed indicate that reaching the predicted topological phase requires electrostatic control of the chemical potential both within the junctions and at the centers of the plaquettes, the present un-gated devices provide an opportunity to explore the underlying collective Josephson physics in this strongly coupled regime. Indeed, we observe a pronounced suppression and subsequent re-emergence of the supercurrent with increasing in-plane magnetic field—a behavior unique to the small super-honeycomb array, where $\xi_{\mathrm{ABS}}>d_{\mathrm{JJ}}$.

In conclusion, we have studied the effect of in-plane and perpendicular magnetic fields on planar Al-InAs JJAs of square and super-honeycomb geometries in the Andreev crystal regime and a control super-honeycomb outside of the Andreev crystal regime. All arrays exhibit critical current features at finite frustration that reflect their respective unit cell areas and stable vortex configurations. By applying an in-plane magnetic field, we observe critical current oscillations as a function of field orientation, relating to the relative angle of the supercurrent in a given JJ and explained by Rashba-dominant spin-orbit interaction. The square and small super-honeycomb geometries at zero frustration show decreasing critical current with increasing in-plane magnetic field. However, at a finite perpendicular field, the behaviors of the three arrays differ. The square array at $f = 1/2$ (S.\ref{supp:fig:square-angle}
(a)) shows a continuous decrease in $I_c$, but for the small super-honeycomb geometry exhibits a distinct minimum in $I_c$ at $f = 1$. The large super-honeycomb array does not exhibit this suppression and re-emergence of supercurrent at $f =1$.  Indicating that this feature may relate to both the length scale of the small super-honeycomb array being within the Andreev crystal regime and the broken symmetry of the vortex lattice at $f =1$. Finally, the numerical calculations predict stable vortex structures and capture in-plane field-induced vortex-lattice disorder. Investigations of junction arrays that break mirror symmetry and are within this $\xi_{\text{ABS}}>d_{\text{JJ}}$ limit are a step towards the 2D topological state as proposed by \cite{lesser2024josephson}. To investigate this phase, future studies could alter the array length scales to explore the isolated junction limit ($\xi_{\text{ABS}}\ll d_{\text{JJ}}$), or leverage the gate-tunability of the InAs 2DEG to vary the chemical potential in the semiconductor regions, the inter-island coupling, and spin-orbit interaction. 




\textit{Acknowledgments} --- This work was supported by ONR N00014-22-1-2764 and ONR N00014-21-1-2450.  This work was supported by the
DFG (CRC/Transregio 183, EI 519/7-1), the Israel Science Foundation ISF (Grant No 1914/24), ISF Quantum Science and Technology (2074/19), and by the European Research Council (ERC) under the European Union's Horizon Europe research and innovation programme (PoC Grant Agreement No. 101287454). We acknowledge fruitful discussions with Felix von Oppen.


\bibliographystyle{apsrev4-2}
\bibliography{bibs/bib}

\begin{thebibliography}{54}%
\makeatletter
\providecommand \@ifxundefined [1]{%
 \@ifx{#1\undefined}
}%
\providecommand \@ifnum [1]{%
 \ifnum #1\expandafter \@firstoftwo
 \else \expandafter \@secondoftwo
 \fi
}%
\providecommand \@ifx [1]{%
 \ifx #1\expandafter \@firstoftwo
 \else \expandafter \@secondoftwo
 \fi
}%
\providecommand \natexlab [1]{#1}%
\providecommand \enquote  [1]{``#1''}%
\providecommand \bibnamefont  [1]{#1}%
\providecommand \bibfnamefont [1]{#1}%
\providecommand \citenamefont [1]{#1}%
\providecommand \href@noop [0]{\@secondoftwo}%
\providecommand \href [0]{\begingroup \@sanitize@url \@href}%
\providecommand \@href[1]{\@@startlink{#1}\@@href}%
\providecommand \@@href[1]{\endgroup#1\@@endlink}%
\providecommand \@sanitize@url [0]{\catcode `\\12\catcode `\$12\catcode `\&12\catcode `\#12\catcode `\^12\catcode `\_12\catcode `\%12\relax}%
\providecommand \@@startlink[1]{}%
\providecommand \@@endlink[0]{}%
\providecommand \url  [0]{\begingroup\@sanitize@url \@url }%
\providecommand \@url [1]{\endgroup\@href {#1}{\urlprefix }}%
\providecommand \urlprefix  [0]{URL }%
\providecommand \Eprint [0]{\href }%
\providecommand \doibase [0]{https://doi.org/}%
\providecommand \selectlanguage [0]{\@gobble}%
\providecommand \bibinfo  [0]{\@secondoftwo}%
\providecommand \bibfield  [0]{\@secondoftwo}%
\providecommand \translation [1]{[#1]}%
\providecommand \BibitemOpen [0]{}%
\providecommand \bibitemStop [0]{}%
\providecommand \bibitemNoStop [0]{.\EOS\space}%
\providecommand \EOS [0]{\spacefactor3000\relax}%
\providecommand \BibitemShut  [1]{\csname bibitem#1\endcsname}%
\let\auto@bib@innerbib\@empty
\bibitem [{\citenamefont {Resnick}\ \emph {et~al.}(1981)\citenamefont {Resnick}, \citenamefont {Garland}, \citenamefont {Boyd}, \citenamefont {Shoemaker},\ and\ \citenamefont {Newrock}}]{resnick1981-BKT}%
  \BibitemOpen
  \bibfield  {author} {\bibinfo {author} {\bibfnamefont {D.}~\bibnamefont {Resnick}}, \bibinfo {author} {\bibfnamefont {J.}~\bibnamefont {Garland}}, \bibinfo {author} {\bibfnamefont {J.}~\bibnamefont {Boyd}}, \bibinfo {author} {\bibfnamefont {S.}~\bibnamefont {Shoemaker}},\ and\ \bibinfo {author} {\bibfnamefont {R.}~\bibnamefont {Newrock}},\ }\href@noop {} {\bibfield  {journal} {\bibinfo  {journal} {Physical Review Letters}\ }\textbf {\bibinfo {volume} {47}},\ \bibinfo {pages} {1542} (\bibinfo {year} {1981})}\BibitemShut {NoStop}%
\bibitem [{\citenamefont {Cosmic}\ \emph {et~al.}(2020)\citenamefont {Cosmic}, \citenamefont {Kawabata}, \citenamefont {Ashida}, \citenamefont {Ikegami}, \citenamefont {Furukawa}, \citenamefont {Patil}, \citenamefont {Taylor},\ and\ \citenamefont {Nakamura}}]{cosmic2020probing-BKT}%
  \BibitemOpen
  \bibfield  {author} {\bibinfo {author} {\bibfnamefont {R.}~\bibnamefont {Cosmic}}, \bibinfo {author} {\bibfnamefont {K.}~\bibnamefont {Kawabata}}, \bibinfo {author} {\bibfnamefont {Y.}~\bibnamefont {Ashida}}, \bibinfo {author} {\bibfnamefont {H.}~\bibnamefont {Ikegami}}, \bibinfo {author} {\bibfnamefont {S.}~\bibnamefont {Furukawa}}, \bibinfo {author} {\bibfnamefont {P.}~\bibnamefont {Patil}}, \bibinfo {author} {\bibfnamefont {J.}~\bibnamefont {Taylor}},\ and\ \bibinfo {author} {\bibfnamefont {Y.}~\bibnamefont {Nakamura}},\ }\href@noop {} {\bibfield  {journal} {\bibinfo  {journal} {Physical Review B}\ }\textbf {\bibinfo {volume} {102}},\ \bibinfo {pages} {094509} (\bibinfo {year} {2020})}\BibitemShut {NoStop}%
\bibitem [{\citenamefont {Van~Wees}\ \emph {et~al.}(1987)\citenamefont {Van~Wees}, \citenamefont {Van~der Zant},\ and\ \citenamefont {Mooij}}]{van1987phase}%
  \BibitemOpen
  \bibfield  {author} {\bibinfo {author} {\bibfnamefont {B.}~\bibnamefont {Van~Wees}}, \bibinfo {author} {\bibfnamefont {H.}~\bibnamefont {Van~der Zant}},\ and\ \bibinfo {author} {\bibfnamefont {J.}~\bibnamefont {Mooij}},\ }\href@noop {} {\bibfield  {journal} {\bibinfo  {journal} {Physical Review B}\ }\textbf {\bibinfo {volume} {35}},\ \bibinfo {pages} {7291} (\bibinfo {year} {1987})}\BibitemShut {NoStop}%
\bibitem [{\citenamefont {Abraham}\ \emph {et~al.}(1982)\citenamefont {Abraham}, \citenamefont {Lobb}, \citenamefont {Tinkham},\ and\ \citenamefont {Klapwijk}}]{abraham1982resistive}%
  \BibitemOpen
  \bibfield  {author} {\bibinfo {author} {\bibfnamefont {D.~W.}\ \bibnamefont {Abraham}}, \bibinfo {author} {\bibfnamefont {C.}~\bibnamefont {Lobb}}, \bibinfo {author} {\bibfnamefont {M.}~\bibnamefont {Tinkham}},\ and\ \bibinfo {author} {\bibfnamefont {T.}~\bibnamefont {Klapwijk}},\ }\href@noop {} {\bibfield  {journal} {\bibinfo  {journal} {Physical Review B}\ }\textbf {\bibinfo {volume} {26}},\ \bibinfo {pages} {5268} (\bibinfo {year} {1982})}\BibitemShut {NoStop}%
\bibitem [{\citenamefont {Martinoli}\ and\ \citenamefont {Leemann}(2000)}]{martinoli2000two}%
  \BibitemOpen
  \bibfield  {author} {\bibinfo {author} {\bibfnamefont {P.}~\bibnamefont {Martinoli}}\ and\ \bibinfo {author} {\bibfnamefont {C.}~\bibnamefont {Leemann}},\ }\href@noop {} {\bibfield  {journal} {\bibinfo  {journal} {Journal of Low Temperature Physics}\ }\textbf {\bibinfo {volume} {118}},\ \bibinfo {pages} {699} (\bibinfo {year} {2000})}\BibitemShut {NoStop}%
\bibitem [{\citenamefont {Newrock}\ \emph {et~al.}(2000)\citenamefont {Newrock}, \citenamefont {Lobb}, \citenamefont {Geigenm{\"u}ller},\ and\ \citenamefont {Octavio}}]{newrock2000two}%
  \BibitemOpen
  \bibfield  {author} {\bibinfo {author} {\bibfnamefont {R.}~\bibnamefont {Newrock}}, \bibinfo {author} {\bibfnamefont {C.}~\bibnamefont {Lobb}}, \bibinfo {author} {\bibfnamefont {U.}~\bibnamefont {Geigenm{\"u}ller}},\ and\ \bibinfo {author} {\bibfnamefont {M.}~\bibnamefont {Octavio}},\ }in\ \href@noop {} {\emph {\bibinfo {booktitle} {Solid state physics}}},\ Vol.~\bibinfo {volume} {54}\ (\bibinfo  {publisher} {Elsevier},\ \bibinfo {year} {2000})\ pp.\ \bibinfo {pages} {263--512}\BibitemShut {NoStop}%
\bibitem [{\citenamefont {Fazio}\ and\ \citenamefont {Van Der~Zant}(2001)}]{fazio2001quantum}%
  \BibitemOpen
  \bibfield  {author} {\bibinfo {author} {\bibfnamefont {R.}~\bibnamefont {Fazio}}\ and\ \bibinfo {author} {\bibfnamefont {H.}~\bibnamefont {Van Der~Zant}},\ }\href@noop {} {\bibfield  {journal} {\bibinfo  {journal} {Physics Reports}\ }\textbf {\bibinfo {volume} {355}},\ \bibinfo {pages} {235} (\bibinfo {year} {2001})}\BibitemShut {NoStop}%
\bibitem [{\citenamefont {B{\o}ttcher}\ \emph {et~al.}(2018)\citenamefont {B{\o}ttcher}, \citenamefont {Nichele}, \citenamefont {Kjaergaard}, \citenamefont {Suominen}, \citenamefont {Shabani}, \citenamefont {Palmstr{\o}m},\ and\ \citenamefont {Marcus}}]{bottcher2018anomolous}%
  \BibitemOpen
  \bibfield  {author} {\bibinfo {author} {\bibfnamefont {C.}~\bibnamefont {B{\o}ttcher}}, \bibinfo {author} {\bibfnamefont {F.}~\bibnamefont {Nichele}}, \bibinfo {author} {\bibfnamefont {M.}~\bibnamefont {Kjaergaard}}, \bibinfo {author} {\bibfnamefont {H.}~\bibnamefont {Suominen}}, \bibinfo {author} {\bibfnamefont {J.}~\bibnamefont {Shabani}}, \bibinfo {author} {\bibfnamefont {C.}~\bibnamefont {Palmstr{\o}m}},\ and\ \bibinfo {author} {\bibfnamefont {C.}~\bibnamefont {Marcus}},\ }\href@noop {} {\bibfield  {journal} {\bibinfo  {journal} {Nature Physics}\ }\textbf {\bibinfo {volume} {14}},\ \bibinfo {pages} {1138} (\bibinfo {year} {2018})}\BibitemShut {NoStop}%
\bibitem [{\citenamefont {Teitel}\ and\ \citenamefont {Jayaprakash}(1983{\natexlab{a}})}]{teitel1983phase-xy}%
  \BibitemOpen
  \bibfield  {author} {\bibinfo {author} {\bibfnamefont {S.}~\bibnamefont {Teitel}}\ and\ \bibinfo {author} {\bibfnamefont {C.}~\bibnamefont {Jayaprakash}},\ }\href@noop {} {\bibfield  {journal} {\bibinfo  {journal} {Physical Review B}\ }\textbf {\bibinfo {volume} {27}},\ \bibinfo {pages} {598} (\bibinfo {year} {1983}{\natexlab{a}})}\BibitemShut {NoStop}%
\bibitem [{\citenamefont {Ashrafuzzaman}\ \emph {et~al.}(2003)\citenamefont {Ashrafuzzaman}, \citenamefont {Capezzali},\ and\ \citenamefont {Beck}}]{ashrafuzzaman2003BKT-XYmodel}%
  \BibitemOpen
  \bibfield  {author} {\bibinfo {author} {\bibfnamefont {M.}~\bibnamefont {Ashrafuzzaman}}, \bibinfo {author} {\bibfnamefont {M.}~\bibnamefont {Capezzali}},\ and\ \bibinfo {author} {\bibfnamefont {H.}~\bibnamefont {Beck}},\ }\href@noop {} {\bibfield  {journal} {\bibinfo  {journal} {Physical Review B}\ }\textbf {\bibinfo {volume} {68}},\ \bibinfo {pages} {052502} (\bibinfo {year} {2003})}\BibitemShut {NoStop}%
\bibitem [{\citenamefont {Van~der Zant}\ \emph {et~al.}(1992)\citenamefont {Van~der Zant}, \citenamefont {Fritschy}, \citenamefont {Elion}, \citenamefont {Geerligs},\ and\ \citenamefont {Mooij}}]{van1992field}%
  \BibitemOpen
  \bibfield  {author} {\bibinfo {author} {\bibfnamefont {H.}~\bibnamefont {Van~der Zant}}, \bibinfo {author} {\bibfnamefont {F.}~\bibnamefont {Fritschy}}, \bibinfo {author} {\bibfnamefont {W.}~\bibnamefont {Elion}}, \bibinfo {author} {\bibfnamefont {L.}~\bibnamefont {Geerligs}},\ and\ \bibinfo {author} {\bibfnamefont {J.}~\bibnamefont {Mooij}},\ }\href@noop {} {\bibfield  {journal} {\bibinfo  {journal} {Physical review letters}\ }\textbf {\bibinfo {volume} {69}},\ \bibinfo {pages} {2971} (\bibinfo {year} {1992})}\BibitemShut {NoStop}%
\bibitem [{\citenamefont {Trias}\ \emph {et~al.}(1995)\citenamefont {Trias}, \citenamefont {Phillips}, \citenamefont {van~der Zant},\ and\ \citenamefont {Orlando}}]{trias1995self}%
  \BibitemOpen
  \bibfield  {author} {\bibinfo {author} {\bibfnamefont {E.}~\bibnamefont {Trias}}, \bibinfo {author} {\bibfnamefont {J.}~\bibnamefont {Phillips}}, \bibinfo {author} {\bibfnamefont {H.}~\bibnamefont {van~der Zant}},\ and\ \bibinfo {author} {\bibfnamefont {T.}~\bibnamefont {Orlando}},\ }\href@noop {} {\bibfield  {journal} {\bibinfo  {journal} {IEEE Transactions on Applied Superconductivity}\ }\textbf {\bibinfo {volume} {5}},\ \bibinfo {pages} {2707} (\bibinfo {year} {1995})}\BibitemShut {NoStop}%
\bibitem [{\citenamefont {B{\o}ttcher}\ \emph {et~al.}(2023)\citenamefont {B{\o}ttcher}, \citenamefont {Nichele}, \citenamefont {Shabani}, \citenamefont {Palmstr{\o}m},\ and\ \citenamefont {Marcus}}]{bottcher2023dynamical}%
  \BibitemOpen
  \bibfield  {author} {\bibinfo {author} {\bibfnamefont {C.}~\bibnamefont {B{\o}ttcher}}, \bibinfo {author} {\bibfnamefont {F.}~\bibnamefont {Nichele}}, \bibinfo {author} {\bibfnamefont {J.}~\bibnamefont {Shabani}}, \bibinfo {author} {\bibfnamefont {C.}~\bibnamefont {Palmstr{\o}m}},\ and\ \bibinfo {author} {\bibfnamefont {C.}~\bibnamefont {Marcus}},\ }\href@noop {} {\bibfield  {journal} {\bibinfo  {journal} {Physical Review B}\ }\textbf {\bibinfo {volume} {108}},\ \bibinfo {pages} {134517} (\bibinfo {year} {2023})}\BibitemShut {NoStop}%
\bibitem [{\citenamefont {B{\o}ttcher}\ \emph {et~al.}(2024)\citenamefont {B{\o}ttcher}, \citenamefont {Nichele}, \citenamefont {Shabani}, \citenamefont {Palmstr{\o}m},\ and\ \citenamefont {Marcus}}]{bottcher2024-BKT}%
  \BibitemOpen
  \bibfield  {author} {\bibinfo {author} {\bibfnamefont {C.}~\bibnamefont {B{\o}ttcher}}, \bibinfo {author} {\bibfnamefont {F.}~\bibnamefont {Nichele}}, \bibinfo {author} {\bibfnamefont {J.}~\bibnamefont {Shabani}}, \bibinfo {author} {\bibfnamefont {C.}~\bibnamefont {Palmstr{\o}m}},\ and\ \bibinfo {author} {\bibfnamefont {C.}~\bibnamefont {Marcus}},\ }\href@noop {} {\bibfield  {journal} {\bibinfo  {journal} {Physical Review B}\ }\textbf {\bibinfo {volume} {110}},\ \bibinfo {pages} {L180502} (\bibinfo {year} {2024})}\BibitemShut {NoStop}%
\bibitem [{\citenamefont {Sasmal}\ \emph {et~al.}(2025)\citenamefont {Sasmal}, \citenamefont {Efthymiou-Tsironi}, \citenamefont {Nagda}, \citenamefont {Fugl}, \citenamefont {Olsen}, \citenamefont {Krizek}, \citenamefont {Marcus},\ and\ \citenamefont {Vaitiek{\.e}nas}}]{sasmal2025voltage-tune-channel-gate}%
  \BibitemOpen
  \bibfield  {author} {\bibinfo {author} {\bibfnamefont {S.}~\bibnamefont {Sasmal}}, \bibinfo {author} {\bibfnamefont {M.}~\bibnamefont {Efthymiou-Tsironi}}, \bibinfo {author} {\bibfnamefont {G.}~\bibnamefont {Nagda}}, \bibinfo {author} {\bibfnamefont {E.}~\bibnamefont {Fugl}}, \bibinfo {author} {\bibfnamefont {L.}~\bibnamefont {Olsen}}, \bibinfo {author} {\bibfnamefont {F.}~\bibnamefont {Krizek}}, \bibinfo {author} {\bibfnamefont {C.}~\bibnamefont {Marcus}},\ and\ \bibinfo {author} {\bibfnamefont {S.}~\bibnamefont {Vaitiek{\.e}nas}},\ }\href@noop {} {\bibfield  {journal} {\bibinfo  {journal} {Physical Review Letters}\ }\textbf {\bibinfo {volume} {135}},\ \bibinfo {pages} {156301} (\bibinfo {year} {2025})}\BibitemShut {NoStop}%
\bibitem [{\citenamefont {Reinhardt}\ \emph {et~al.}(2025)\citenamefont {Reinhardt}, \citenamefont {Penner}, \citenamefont {Berger}, \citenamefont {Baumgartner}, \citenamefont {Gronin}, \citenamefont {Gardner}, \citenamefont {Lindemann}, \citenamefont {Manfra}, \citenamefont {Glazman}, \citenamefont {von Oppen} \emph {et~al.}}]{reinhardt2025-diode-array}%
  \BibitemOpen
  \bibfield  {author} {\bibinfo {author} {\bibfnamefont {S.}~\bibnamefont {Reinhardt}}, \bibinfo {author} {\bibfnamefont {A.-G.}\ \bibnamefont {Penner}}, \bibinfo {author} {\bibfnamefont {J.}~\bibnamefont {Berger}}, \bibinfo {author} {\bibfnamefont {C.}~\bibnamefont {Baumgartner}}, \bibinfo {author} {\bibfnamefont {S.}~\bibnamefont {Gronin}}, \bibinfo {author} {\bibfnamefont {G.~C.}\ \bibnamefont {Gardner}}, \bibinfo {author} {\bibfnamefont {T.}~\bibnamefont {Lindemann}}, \bibinfo {author} {\bibfnamefont {M.~J.}\ \bibnamefont {Manfra}}, \bibinfo {author} {\bibfnamefont {L.~I.}\ \bibnamefont {Glazman}}, \bibinfo {author} {\bibfnamefont {F.}~\bibnamefont {von Oppen}}, \emph {et~al.},\ }\href@noop {} {\bibfield  {journal} {\bibinfo  {journal} {Physical Review B}\ }\textbf {\bibinfo {volume} {112}},\ \bibinfo {pages} {224503} (\bibinfo {year} {2025})}\BibitemShut {NoStop}%
\bibitem [{\citenamefont {Baumgartner}\ \emph {et~al.}(2022)\citenamefont {Baumgartner}, \citenamefont {Fuchs}, \citenamefont {Costa}, \citenamefont {Reinhardt}, \citenamefont {Gronin}, \citenamefont {Gardner}, \citenamefont {Lindemann}, \citenamefont {Manfra}, \citenamefont {Faria~Junior}, \citenamefont {Kochan} \emph {et~al.}}]{baumgartner2021-1DJJA-B_par}%
  \BibitemOpen
  \bibfield  {author} {\bibinfo {author} {\bibfnamefont {C.}~\bibnamefont {Baumgartner}}, \bibinfo {author} {\bibfnamefont {L.}~\bibnamefont {Fuchs}}, \bibinfo {author} {\bibfnamefont {A.}~\bibnamefont {Costa}}, \bibinfo {author} {\bibfnamefont {S.}~\bibnamefont {Reinhardt}}, \bibinfo {author} {\bibfnamefont {S.}~\bibnamefont {Gronin}}, \bibinfo {author} {\bibfnamefont {G.~C.}\ \bibnamefont {Gardner}}, \bibinfo {author} {\bibfnamefont {T.}~\bibnamefont {Lindemann}}, \bibinfo {author} {\bibfnamefont {M.~J.}\ \bibnamefont {Manfra}}, \bibinfo {author} {\bibfnamefont {P.~E.}\ \bibnamefont {Faria~Junior}}, \bibinfo {author} {\bibfnamefont {D.}~\bibnamefont {Kochan}}, \emph {et~al.},\ }\href@noop {} {\bibfield  {journal} {\bibinfo  {journal} {Nature nanotechnology}\ }\textbf {\bibinfo {volume} {17}},\ \bibinfo {pages} {39} (\bibinfo {year} {2022})}\BibitemShut {NoStop}%
\bibitem [{\citenamefont {Teitel}\ and\ \citenamefont {Jayaprakash}(1983{\natexlab{b}})}]{teitel-jayprakash1983}%
  \BibitemOpen
  \bibfield  {author} {\bibinfo {author} {\bibfnamefont {S.}~\bibnamefont {Teitel}}\ and\ \bibinfo {author} {\bibfnamefont {C.}~\bibnamefont {Jayaprakash}},\ }\href@noop {} {\bibfield  {journal} {\bibinfo  {journal} {Physical review letters}\ }\textbf {\bibinfo {volume} {51}},\ \bibinfo {pages} {1999} (\bibinfo {year} {1983}{\natexlab{b}})}\BibitemShut {NoStop}%
\bibitem [{\citenamefont {Tinkham}\ \emph {et~al.}(1983)\citenamefont {Tinkham}, \citenamefont {Abraham},\ and\ \citenamefont {Lobb}}]{tinkham1983periodic-vortex}%
  \BibitemOpen
  \bibfield  {author} {\bibinfo {author} {\bibfnamefont {M.}~\bibnamefont {Tinkham}}, \bibinfo {author} {\bibfnamefont {D.~W.}\ \bibnamefont {Abraham}},\ and\ \bibinfo {author} {\bibfnamefont {C.}~\bibnamefont {Lobb}},\ }\href@noop {} {\bibfield  {journal} {\bibinfo  {journal} {Physical Review B}\ }\textbf {\bibinfo {volume} {28}},\ \bibinfo {pages} {6578} (\bibinfo {year} {1983})}\BibitemShut {NoStop}%
\bibitem [{\citenamefont {Halsey}(1985)}]{halsey1985-square-lattice-theory}%
  \BibitemOpen
  \bibfield  {author} {\bibinfo {author} {\bibfnamefont {T.~C.}\ \bibnamefont {Halsey}},\ }\href@noop {} {\bibfield  {journal} {\bibinfo  {journal} {Physical Review B}\ }\textbf {\bibinfo {volume} {31}},\ \bibinfo {pages} {5728} (\bibinfo {year} {1985})}\BibitemShut {NoStop}%
\bibitem [{\citenamefont {Lin}\ and\ \citenamefont {Nori}(2002)}]{lin2002-varyingJJA-patterns}%
  \BibitemOpen
  \bibfield  {author} {\bibinfo {author} {\bibfnamefont {Y.-L.}\ \bibnamefont {Lin}}\ and\ \bibinfo {author} {\bibfnamefont {F.}~\bibnamefont {Nori}},\ }\href@noop {} {\bibfield  {journal} {\bibinfo  {journal} {Physical Review B}\ }\textbf {\bibinfo {volume} {65}},\ \bibinfo {pages} {214504} (\bibinfo {year} {2002})}\BibitemShut {NoStop}%
\bibitem [{\citenamefont {Park}\ and\ \citenamefont {Huse}(2001)}]{park2001-kagome-JJA-theory}%
  \BibitemOpen
  \bibfield  {author} {\bibinfo {author} {\bibfnamefont {K.}~\bibnamefont {Park}}\ and\ \bibinfo {author} {\bibfnamefont {D.~A.}\ \bibnamefont {Huse}},\ }\href@noop {} {\bibfield  {journal} {\bibinfo  {journal} {Physical Review B}\ }\textbf {\bibinfo {volume} {64}},\ \bibinfo {pages} {134522} (\bibinfo {year} {2001})}\BibitemShut {NoStop}%
\bibitem [{\citenamefont {Teller}\ \emph {et~al.}(2025)\citenamefont {Teller}, \citenamefont {Sch{\"a}fer}, \citenamefont {Moors}, \citenamefont {Bennemann}, \citenamefont {Lyatti}, \citenamefont {Lentz}, \citenamefont {Gr{\"u}tzmacher}, \citenamefont {Riwar},\ and\ \citenamefont {Sch{\"a}pers}}]{teller2025frustrated-frustration}%
  \BibitemOpen
  \bibfield  {author} {\bibinfo {author} {\bibfnamefont {J.}~\bibnamefont {Teller}}, \bibinfo {author} {\bibfnamefont {C.}~\bibnamefont {Sch{\"a}fer}}, \bibinfo {author} {\bibfnamefont {K.}~\bibnamefont {Moors}}, \bibinfo {author} {\bibfnamefont {B.}~\bibnamefont {Bennemann}}, \bibinfo {author} {\bibfnamefont {M.}~\bibnamefont {Lyatti}}, \bibinfo {author} {\bibfnamefont {F.}~\bibnamefont {Lentz}}, \bibinfo {author} {\bibfnamefont {D.}~\bibnamefont {Gr{\"u}tzmacher}}, \bibinfo {author} {\bibfnamefont {R.-P.}\ \bibnamefont {Riwar}},\ and\ \bibinfo {author} {\bibfnamefont {T.}~\bibnamefont {Sch{\"a}pers}},\ }\href@noop {} {\bibfield  {journal} {\bibinfo  {journal} {Physical review letters}\ }\textbf {\bibinfo {volume} {135}},\ \bibinfo {pages} {156002} (\bibinfo {year} {2025})}\BibitemShut {NoStop}%
\bibitem [{\citenamefont {Bondar}\ \emph {et~al.}(2025)\citenamefont {Bondar}, \citenamefont {Banszerus}, \citenamefont {Marshall}, \citenamefont {Lindemann}, \citenamefont {Zhang}, \citenamefont {Manfra}, \citenamefont {Marcus},\ and\ \citenamefont {Vaitiek{\.e}nas}}]{bondar2025-dice}%
  \BibitemOpen
  \bibfield  {author} {\bibinfo {author} {\bibfnamefont {J.}~\bibnamefont {Bondar}}, \bibinfo {author} {\bibfnamefont {L.}~\bibnamefont {Banszerus}}, \bibinfo {author} {\bibfnamefont {W.}~\bibnamefont {Marshall}}, \bibinfo {author} {\bibfnamefont {T.}~\bibnamefont {Lindemann}}, \bibinfo {author} {\bibfnamefont {T.}~\bibnamefont {Zhang}}, \bibinfo {author} {\bibfnamefont {M.}~\bibnamefont {Manfra}}, \bibinfo {author} {\bibfnamefont {C.}~\bibnamefont {Marcus}},\ and\ \bibinfo {author} {\bibfnamefont {S.}~\bibnamefont {Vaitiek{\.e}nas}},\ }\href@noop {} {\bibfield  {journal} {\bibinfo  {journal} {arXiv preprint arXiv:2510.07412}\ } (\bibinfo {year} {2025})}\BibitemShut {NoStop}%
\bibitem [{\citenamefont {Tinkham}(2004)}]{tinkham2004introduction}%
  \BibitemOpen
  \bibfield  {author} {\bibinfo {author} {\bibfnamefont {M.}~\bibnamefont {Tinkham}},\ }\href@noop {} {\emph {\bibinfo {title} {Introduction to superconductivity}}}\ (\bibinfo  {publisher} {Courier Corporation},\ \bibinfo {year} {2004})\BibitemShut {NoStop}%
\bibitem [{\citenamefont {Andreev}\ \emph {et~al.}(1965)\citenamefont {Andreev} \emph {et~al.}}]{andreev1965thermal-original-andreev-reflection}%
  \BibitemOpen
  \bibfield  {author} {\bibinfo {author} {\bibfnamefont {A.}~\bibnamefont {Andreev}} \emph {et~al.},\ }\href@noop {} {\bibfield  {journal} {\bibinfo  {journal} {Sov. Phys. JETP}\ }\textbf {\bibinfo {volume} {20}},\ \bibinfo {pages} {1490} (\bibinfo {year} {1965})}\BibitemShut {NoStop}%
\bibitem [{\citenamefont {Blonder}\ \emph {et~al.}(1982)\citenamefont {Blonder}, \citenamefont {Tinkham},\ and\ \citenamefont {Klapwijk}}]{blonder1982-ABS}%
  \BibitemOpen
  \bibfield  {author} {\bibinfo {author} {\bibfnamefont {G.}~\bibnamefont {Blonder}}, \bibinfo {author} {\bibfnamefont {m.~M.}\ \bibnamefont {Tinkham}},\ and\ \bibinfo {author} {\bibfnamefont {T.}~\bibnamefont {Klapwijk}},\ }\href@noop {} {\bibfield  {journal} {\bibinfo  {journal} {Physical Review B}\ }\textbf {\bibinfo {volume} {25}},\ \bibinfo {pages} {4515} (\bibinfo {year} {1982})}\BibitemShut {NoStop}%
\bibitem [{\citenamefont {Prada}\ \emph {et~al.}(2020)\citenamefont {Prada}, \citenamefont {San-Jose}, \citenamefont {de~Moor}, \citenamefont {Geresdi}, \citenamefont {Lee}, \citenamefont {Klinovaja}, \citenamefont {Loss}, \citenamefont {Nyg{\aa}rd}, \citenamefont {Aguado},\ and\ \citenamefont {Kouwenhoven}}]{prada2020andreev-abs-coherence-length}%
  \BibitemOpen
  \bibfield  {author} {\bibinfo {author} {\bibfnamefont {E.}~\bibnamefont {Prada}}, \bibinfo {author} {\bibfnamefont {P.}~\bibnamefont {San-Jose}}, \bibinfo {author} {\bibfnamefont {M.~W.}\ \bibnamefont {de~Moor}}, \bibinfo {author} {\bibfnamefont {A.}~\bibnamefont {Geresdi}}, \bibinfo {author} {\bibfnamefont {E.~J.}\ \bibnamefont {Lee}}, \bibinfo {author} {\bibfnamefont {J.}~\bibnamefont {Klinovaja}}, \bibinfo {author} {\bibfnamefont {D.}~\bibnamefont {Loss}}, \bibinfo {author} {\bibfnamefont {J.}~\bibnamefont {Nyg{\aa}rd}}, \bibinfo {author} {\bibfnamefont {R.}~\bibnamefont {Aguado}},\ and\ \bibinfo {author} {\bibfnamefont {L.~P.}\ \bibnamefont {Kouwenhoven}},\ }\href@noop {} {\bibfield  {journal} {\bibinfo  {journal} {Nature Reviews Physics}\ }\textbf {\bibinfo {volume} {2}},\ \bibinfo {pages} {575} (\bibinfo {year} {2020})}\BibitemShut {NoStop}%
\bibitem [{\citenamefont {Pillet}\ \emph {et~al.}(2019)\citenamefont {Pillet}, \citenamefont {Benzoni}, \citenamefont {Griesmar}, \citenamefont {Smirr},\ and\ \citenamefont {Girit}}]{pillet2019nonlocal-Andreev-molecules}%
  \BibitemOpen
  \bibfield  {author} {\bibinfo {author} {\bibfnamefont {J.-D.}\ \bibnamefont {Pillet}}, \bibinfo {author} {\bibfnamefont {V.}~\bibnamefont {Benzoni}}, \bibinfo {author} {\bibfnamefont {J.}~\bibnamefont {Griesmar}}, \bibinfo {author} {\bibfnamefont {J.-L.}\ \bibnamefont {Smirr}},\ and\ \bibinfo {author} {\bibfnamefont {C.~O.}\ \bibnamefont {Girit}},\ }\href@noop {} {\bibfield  {journal} {\bibinfo  {journal} {Nano letters}\ }\textbf {\bibinfo {volume} {19}},\ \bibinfo {pages} {7138} (\bibinfo {year} {2019})}\BibitemShut {NoStop}%
\bibitem [{\citenamefont {Pillet}\ \emph {et~al.}(2020)\citenamefont {Pillet}, \citenamefont {Benzoni}, \citenamefont {Griesmar}, \citenamefont {Smirr},\ and\ \citenamefont {Girit}}]{pillet2020scattering-andreev-molecules}%
  \BibitemOpen
  \bibfield  {author} {\bibinfo {author} {\bibfnamefont {J.-D.}\ \bibnamefont {Pillet}}, \bibinfo {author} {\bibfnamefont {V.}~\bibnamefont {Benzoni}}, \bibinfo {author} {\bibfnamefont {J.}~\bibnamefont {Griesmar}}, \bibinfo {author} {\bibfnamefont {J.-L.}\ \bibnamefont {Smirr}},\ and\ \bibinfo {author} {\bibfnamefont {{\c{C}}.}~\bibnamefont {Girit}},\ }\href@noop {} {\bibfield  {journal} {\bibinfo  {journal} {SciPost Physics Core}\ }\textbf {\bibinfo {volume} {2}},\ \bibinfo {pages} {009} (\bibinfo {year} {2020})}\BibitemShut {NoStop}%
\bibitem [{\citenamefont {Kornich}\ \emph {et~al.}(2019)\citenamefont {Kornich}, \citenamefont {Barakov},\ and\ \citenamefont {Nazarov}}]{Kornich-overlapping-ABS-andreev-molecules}%
  \BibitemOpen
  \bibfield  {author} {\bibinfo {author} {\bibfnamefont {V.}~\bibnamefont {Kornich}}, \bibinfo {author} {\bibfnamefont {H.~S.}\ \bibnamefont {Barakov}},\ and\ \bibinfo {author} {\bibfnamefont {Y.~V.}\ \bibnamefont {Nazarov}},\ }\href {https://doi.org/10.1103/PhysRevResearch.1.033004} {\bibfield  {journal} {\bibinfo  {journal} {Phys. Rev. Res.}\ }\textbf {\bibinfo {volume} {1}},\ \bibinfo {pages} {033004} (\bibinfo {year} {2019})}\BibitemShut {NoStop}%
\bibitem [{\citenamefont {Haxell}\ \emph {et~al.}(2023)\citenamefont {Haxell}, \citenamefont {Coraiola}, \citenamefont {Hinderling}, \citenamefont {Ten~Kate}, \citenamefont {Sabonis}, \citenamefont {Svetogorov}, \citenamefont {Belzig}, \citenamefont {Cheah}, \citenamefont {Krizek}, \citenamefont {Schott} \emph {et~al.}}]{haxell2023demonstration-andreev-molecules-exp}%
  \BibitemOpen
  \bibfield  {author} {\bibinfo {author} {\bibfnamefont {D.~Z.}\ \bibnamefont {Haxell}}, \bibinfo {author} {\bibfnamefont {M.}~\bibnamefont {Coraiola}}, \bibinfo {author} {\bibfnamefont {M.}~\bibnamefont {Hinderling}}, \bibinfo {author} {\bibfnamefont {S.~C.}\ \bibnamefont {Ten~Kate}}, \bibinfo {author} {\bibfnamefont {D.}~\bibnamefont {Sabonis}}, \bibinfo {author} {\bibfnamefont {A.~E.}\ \bibnamefont {Svetogorov}}, \bibinfo {author} {\bibfnamefont {W.}~\bibnamefont {Belzig}}, \bibinfo {author} {\bibfnamefont {E.}~\bibnamefont {Cheah}}, \bibinfo {author} {\bibfnamefont {F.}~\bibnamefont {Krizek}}, \bibinfo {author} {\bibfnamefont {R.}~\bibnamefont {Schott}}, \emph {et~al.},\ }\href@noop {} {\bibfield  {journal} {\bibinfo  {journal} {Nano Letters}\ }\textbf {\bibinfo {volume} {23}},\ \bibinfo {pages} {7532} (\bibinfo {year} {2023})}\BibitemShut {NoStop}%
\bibitem [{\citenamefont {Matsuo}\ \emph {et~al.}(2023)\citenamefont {Matsuo}, \citenamefont {Imoto}, \citenamefont {Yokoyama}, \citenamefont {Sato}, \citenamefont {Lindemann}, \citenamefont {Gronin}, \citenamefont {Gardner}, \citenamefont {Nakosai}, \citenamefont {Tanaka}, \citenamefont {Manfra} \emph {et~al.}}]{matsuo2023phase-andreev-molecule-exp}%
  \BibitemOpen
  \bibfield  {author} {\bibinfo {author} {\bibfnamefont {S.}~\bibnamefont {Matsuo}}, \bibinfo {author} {\bibfnamefont {T.}~\bibnamefont {Imoto}}, \bibinfo {author} {\bibfnamefont {T.}~\bibnamefont {Yokoyama}}, \bibinfo {author} {\bibfnamefont {Y.}~\bibnamefont {Sato}}, \bibinfo {author} {\bibfnamefont {T.}~\bibnamefont {Lindemann}}, \bibinfo {author} {\bibfnamefont {S.}~\bibnamefont {Gronin}}, \bibinfo {author} {\bibfnamefont {G.~C.}\ \bibnamefont {Gardner}}, \bibinfo {author} {\bibfnamefont {S.}~\bibnamefont {Nakosai}}, \bibinfo {author} {\bibfnamefont {Y.}~\bibnamefont {Tanaka}}, \bibinfo {author} {\bibfnamefont {M.~J.}\ \bibnamefont {Manfra}}, \emph {et~al.},\ }\href@noop {} {\bibfield  {journal} {\bibinfo  {journal} {Nature Communications}\ }\textbf {\bibinfo {volume} {14}},\ \bibinfo {pages} {8271} (\bibinfo {year} {2023})}\BibitemShut {NoStop}%
\bibitem [{\citenamefont {Dahl}\ \emph {et~al.}(2025)\citenamefont {Dahl}, \citenamefont {Maiani}, \citenamefont {Geier}, \citenamefont {Shabani},\ and\ \citenamefont {Flensberg}}]{dahl2025andreev-crystal}%
  \BibitemOpen
  \bibfield  {author} {\bibinfo {author} {\bibfnamefont {A.~E.}\ \bibnamefont {Dahl}}, \bibinfo {author} {\bibfnamefont {A.}~\bibnamefont {Maiani}}, \bibinfo {author} {\bibfnamefont {M.}~\bibnamefont {Geier}}, \bibinfo {author} {\bibfnamefont {J.}~\bibnamefont {Shabani}},\ and\ \bibinfo {author} {\bibfnamefont {K.}~\bibnamefont {Flensberg}},\ }\href@noop {} {\bibfield  {journal} {\bibinfo  {journal} {arXiv preprint arXiv:2508.11768}\ } (\bibinfo {year} {2025})}\BibitemShut {NoStop}%
\bibitem [{\citenamefont {Shabani}\ \emph {et~al.}(2016)\citenamefont {Shabani}, \citenamefont {Kj{\ae}rgaard}, \citenamefont {Suominen}, \citenamefont {Kim}, \citenamefont {Nichele}, \citenamefont {Pakrouski}, \citenamefont {Stankevic}, \citenamefont {Lutchyn}, \citenamefont {Krogstrup}, \citenamefont {Feidenhans'l} \emph {et~al.}}]{shabani2016AlInAs}%
  \BibitemOpen
  \bibfield  {author} {\bibinfo {author} {\bibfnamefont {J.}~\bibnamefont {Shabani}}, \bibinfo {author} {\bibfnamefont {M.}~\bibnamefont {Kj{\ae}rgaard}}, \bibinfo {author} {\bibfnamefont {H.~J.}\ \bibnamefont {Suominen}}, \bibinfo {author} {\bibfnamefont {Y.}~\bibnamefont {Kim}}, \bibinfo {author} {\bibfnamefont {F.}~\bibnamefont {Nichele}}, \bibinfo {author} {\bibfnamefont {K.}~\bibnamefont {Pakrouski}}, \bibinfo {author} {\bibfnamefont {T.}~\bibnamefont {Stankevic}}, \bibinfo {author} {\bibfnamefont {R.~M.}\ \bibnamefont {Lutchyn}}, \bibinfo {author} {\bibfnamefont {P.}~\bibnamefont {Krogstrup}}, \bibinfo {author} {\bibfnamefont {R.}~\bibnamefont {Feidenhans'l}}, \emph {et~al.},\ }\href@noop {} {\bibfield  {journal} {\bibinfo  {journal} {Physical Review B}\ }\textbf {\bibinfo {volume} {93}},\ \bibinfo {pages} {155402} (\bibinfo {year} {2016})}\BibitemShut {NoStop}%
\bibitem [{\citenamefont {Farzaneh}\ \emph {et~al.}(2024)\citenamefont {Farzaneh}, \citenamefont {Hatefipour}, \citenamefont {Schiela}, \citenamefont {Lotfizadeh}, \citenamefont {Yu}, \citenamefont {Elfeky}, \citenamefont {Strickland}, \citenamefont {Matos-Abiague},\ and\ \citenamefont {Shabani}}]{farzaneh2024observing}%
  \BibitemOpen
  \bibfield  {author} {\bibinfo {author} {\bibfnamefont {S.}~\bibnamefont {Farzaneh}}, \bibinfo {author} {\bibfnamefont {M.}~\bibnamefont {Hatefipour}}, \bibinfo {author} {\bibfnamefont {W.~F.}\ \bibnamefont {Schiela}}, \bibinfo {author} {\bibfnamefont {N.}~\bibnamefont {Lotfizadeh}}, \bibinfo {author} {\bibfnamefont {P.}~\bibnamefont {Yu}}, \bibinfo {author} {\bibfnamefont {B.~H.}\ \bibnamefont {Elfeky}}, \bibinfo {author} {\bibfnamefont {W.~M.}\ \bibnamefont {Strickland}}, \bibinfo {author} {\bibfnamefont {A.}~\bibnamefont {Matos-Abiague}},\ and\ \bibinfo {author} {\bibfnamefont {J.}~\bibnamefont {Shabani}},\ }\href@noop {} {\bibfield  {journal} {\bibinfo  {journal} {Physical Review Research}\ }\textbf {\bibinfo {volume} {6}},\ \bibinfo {pages} {013039} (\bibinfo {year} {2024})}\BibitemShut {NoStop}%
\bibitem [{\citenamefont {Lesser}\ \emph {et~al.}(2024)\citenamefont {Lesser}, \citenamefont {Stern},\ and\ \citenamefont {Oreg}}]{lesser2024josephson}%
  \BibitemOpen
  \bibfield  {author} {\bibinfo {author} {\bibfnamefont {O.}~\bibnamefont {Lesser}}, \bibinfo {author} {\bibfnamefont {A.}~\bibnamefont {Stern}},\ and\ \bibinfo {author} {\bibfnamefont {Y.}~\bibnamefont {Oreg}},\ }\href@noop {} {\bibfield  {journal} {\bibinfo  {journal} {Physical Review B}\ }\textbf {\bibinfo {volume} {109}},\ \bibinfo {pages} {144519} (\bibinfo {year} {2024})}\BibitemShut {NoStop}%
\bibitem [{\citenamefont {Wickramasinghe}\ \emph {et~al.}(2018)\citenamefont {Wickramasinghe}, \citenamefont {Mayer}, \citenamefont {Yuan}, \citenamefont {Nguyen}, \citenamefont {Jiao}, \citenamefont {Manucharyan},\ and\ \citenamefont {Shabani}}]{wickramasinghe2018transport-AlInAs}%
  \BibitemOpen
  \bibfield  {author} {\bibinfo {author} {\bibfnamefont {K.~S.}\ \bibnamefont {Wickramasinghe}}, \bibinfo {author} {\bibfnamefont {W.}~\bibnamefont {Mayer}}, \bibinfo {author} {\bibfnamefont {J.}~\bibnamefont {Yuan}}, \bibinfo {author} {\bibfnamefont {T.}~\bibnamefont {Nguyen}}, \bibinfo {author} {\bibfnamefont {L.}~\bibnamefont {Jiao}}, \bibinfo {author} {\bibfnamefont {V.}~\bibnamefont {Manucharyan}},\ and\ \bibinfo {author} {\bibfnamefont {J.}~\bibnamefont {Shabani}},\ }\href@noop {} {\bibfield  {journal} {\bibinfo  {journal} {Applied Physics Letters}\ }\textbf {\bibinfo {volume} {113}} (\bibinfo {year} {2018})}\BibitemShut {NoStop}%
\bibitem [{\citenamefont {Adachi}(2009)}]{adachi2009-InAs-eff-band-mass}%
  \BibitemOpen
  \bibfield  {author} {\bibinfo {author} {\bibfnamefont {S.}~\bibnamefont {Adachi}},\ }\href@noop {} {\emph {\bibinfo {title} {Properties of semiconductor alloys: group-IV, III-V and II-VI semiconductors}}}\ (\bibinfo  {publisher} {John Wiley \& Sons},\ \bibinfo {year} {2009})\BibitemShut {NoStop}%
\bibitem [{\citenamefont {Schiela}\ \emph {et~al.}(2025{\natexlab{a}})\citenamefont {Schiela}, \citenamefont {Mikalsen}, \citenamefont {Strickland},\ and\ \citenamefont {Shabani}}]{schiela2025geometric}%
  \BibitemOpen
  \bibfield  {author} {\bibinfo {author} {\bibfnamefont {W.~F.}\ \bibnamefont {Schiela}}, \bibinfo {author} {\bibfnamefont {M.}~\bibnamefont {Mikalsen}}, \bibinfo {author} {\bibfnamefont {W.~M.}\ \bibnamefont {Strickland}},\ and\ \bibinfo {author} {\bibfnamefont {J.}~\bibnamefont {Shabani}},\ }\href@noop {} {\bibfield  {journal} {\bibinfo  {journal} {arXiv preprint arXiv:2502.12400}\ } (\bibinfo {year} {2025}{\natexlab{a}})}\BibitemShut {NoStop}%
\bibitem [{\citenamefont {Mayer}\ \emph {et~al.}(2019)\citenamefont {Mayer}, \citenamefont {Yuan}, \citenamefont {Wickramasinghe}, \citenamefont {Nguyen}, \citenamefont {Dartiailh},\ and\ \citenamefont {Shabani}}]{mayer2019abs-coherence-length}%
  \BibitemOpen
  \bibfield  {author} {\bibinfo {author} {\bibfnamefont {W.}~\bibnamefont {Mayer}}, \bibinfo {author} {\bibfnamefont {J.}~\bibnamefont {Yuan}}, \bibinfo {author} {\bibfnamefont {K.~S.}\ \bibnamefont {Wickramasinghe}}, \bibinfo {author} {\bibfnamefont {T.}~\bibnamefont {Nguyen}}, \bibinfo {author} {\bibfnamefont {M.~C.}\ \bibnamefont {Dartiailh}},\ and\ \bibinfo {author} {\bibfnamefont {J.}~\bibnamefont {Shabani}},\ }\href@noop {} {\bibfield  {journal} {\bibinfo  {journal} {Applied Physics Letters}\ }\textbf {\bibinfo {volume} {114}} (\bibinfo {year} {2019})}\BibitemShut {NoStop}%
\bibitem [{\citenamefont {Lankhorst}\ \emph {et~al.}(2018)\citenamefont {Lankhorst}, \citenamefont {Brinkman}, \citenamefont {Hilgenkamp}, \citenamefont {Poccia},\ and\ \citenamefont {Golubov}}]{lankhorst2018annealed}%
  \BibitemOpen
  \bibfield  {author} {\bibinfo {author} {\bibfnamefont {M.}~\bibnamefont {Lankhorst}}, \bibinfo {author} {\bibfnamefont {A.}~\bibnamefont {Brinkman}}, \bibinfo {author} {\bibfnamefont {H.}~\bibnamefont {Hilgenkamp}}, \bibinfo {author} {\bibfnamefont {N.}~\bibnamefont {Poccia}},\ and\ \bibinfo {author} {\bibfnamefont {A.}~\bibnamefont {Golubov}},\ }\href@noop {} {\bibfield  {journal} {\bibinfo  {journal} {Condensed Matter}\ }\textbf {\bibinfo {volume} {3}},\ \bibinfo {pages} {19} (\bibinfo {year} {2018})}\BibitemShut {NoStop}%
\bibitem [{\citenamefont {Rzchowski}\ \emph {et~al.}(1990)\citenamefont {Rzchowski}, \citenamefont {Benz}, \citenamefont {Tinkham},\ and\ \citenamefont {Lobb}}]{rzchowski1990vortex-pinning-potential}%
  \BibitemOpen
  \bibfield  {author} {\bibinfo {author} {\bibfnamefont {M.}~\bibnamefont {Rzchowski}}, \bibinfo {author} {\bibfnamefont {S.}~\bibnamefont {Benz}}, \bibinfo {author} {\bibfnamefont {M.}~\bibnamefont {Tinkham}},\ and\ \bibinfo {author} {\bibfnamefont {C.}~\bibnamefont {Lobb}},\ }\href@noop {} {\bibfield  {journal} {\bibinfo  {journal} {Physical Review B}\ }\textbf {\bibinfo {volume} {42}},\ \bibinfo {pages} {2041} (\bibinfo {year} {1990})}\BibitemShut {NoStop}%
\bibitem [{\citenamefont {Fu}\ and\ \citenamefont {Kane}(2008)}]{fu2008superconducting}%
  \BibitemOpen
  \bibfield  {author} {\bibinfo {author} {\bibfnamefont {L.}~\bibnamefont {Fu}}\ and\ \bibinfo {author} {\bibfnamefont {C.~L.}\ \bibnamefont {Kane}},\ }\href@noop {} {\bibfield  {journal} {\bibinfo  {journal} {Physical review letters}\ }\textbf {\bibinfo {volume} {100}},\ \bibinfo {pages} {096407} (\bibinfo {year} {2008})}\BibitemShut {NoStop}%
\bibitem [{\citenamefont {Pientka}\ \emph {et~al.}(2017{\natexlab{a}})\citenamefont {Pientka}, \citenamefont {Keselman}, \citenamefont {Berg}, \citenamefont {Yacoby}, \citenamefont {Stern},\ and\ \citenamefont {Halperin}}]{pientka2017topological}%
  \BibitemOpen
  \bibfield  {author} {\bibinfo {author} {\bibfnamefont {F.}~\bibnamefont {Pientka}}, \bibinfo {author} {\bibfnamefont {A.}~\bibnamefont {Keselman}}, \bibinfo {author} {\bibfnamefont {E.}~\bibnamefont {Berg}}, \bibinfo {author} {\bibfnamefont {A.}~\bibnamefont {Yacoby}}, \bibinfo {author} {\bibfnamefont {A.}~\bibnamefont {Stern}},\ and\ \bibinfo {author} {\bibfnamefont {B.~I.}\ \bibnamefont {Halperin}},\ }\href@noop {} {\bibfield  {journal} {\bibinfo  {journal} {Physical Review X}\ }\textbf {\bibinfo {volume} {7}},\ \bibinfo {pages} {021032} (\bibinfo {year} {2017}{\natexlab{a}})}\BibitemShut {NoStop}%
\bibitem [{\citenamefont {Schiela}\ \emph {et~al.}(2025{\natexlab{b}})\citenamefont {Schiela}, \citenamefont {Mikalsen}, \citenamefont {Crawford}, \citenamefont {Ili{\'c}}, \citenamefont {Strickland}, \citenamefont {Bergeret},\ and\ \citenamefont {Shabani}}]{schiela2025gate-3-fold-rotation-symmetry}%
  \BibitemOpen
  \bibfield  {author} {\bibinfo {author} {\bibfnamefont {W.~F.}\ \bibnamefont {Schiela}}, \bibinfo {author} {\bibfnamefont {M.}~\bibnamefont {Mikalsen}}, \bibinfo {author} {\bibfnamefont {D.}~\bibnamefont {Crawford}}, \bibinfo {author} {\bibfnamefont {S.}~\bibnamefont {Ili{\'c}}}, \bibinfo {author} {\bibfnamefont {W.~M.}\ \bibnamefont {Strickland}}, \bibinfo {author} {\bibfnamefont {F.~S.}\ \bibnamefont {Bergeret}},\ and\ \bibinfo {author} {\bibfnamefont {J.}~\bibnamefont {Shabani}},\ }\href@noop {} {\bibfield  {journal} {\bibinfo  {journal} {arXiv preprint arXiv:2504.21470}\ } (\bibinfo {year} {2025}{\natexlab{b}})}\BibitemShut {NoStop}%
\bibitem [{\citenamefont {Buzdin}(2008)}]{buzdin2008direct}%
  \BibitemOpen
  \bibfield  {author} {\bibinfo {author} {\bibfnamefont {A.}~\bibnamefont {Buzdin}},\ }\href@noop {} {\bibfield  {journal} {\bibinfo  {journal} {Physical review letters}\ }\textbf {\bibinfo {volume} {101}},\ \bibinfo {pages} {107005} (\bibinfo {year} {2008})}\BibitemShut {NoStop}%
\bibitem [{\citenamefont {Oreg}\ \emph {et~al.}(2010)\citenamefont {Oreg}, \citenamefont {Refael},\ and\ \citenamefont {von Oppen}}]{oreg2010majorana}%
  \BibitemOpen
  \bibfield  {author} {\bibinfo {author} {\bibfnamefont {Y.}~\bibnamefont {Oreg}}, \bibinfo {author} {\bibfnamefont {G.}~\bibnamefont {Refael}},\ and\ \bibinfo {author} {\bibfnamefont {F.}~\bibnamefont {von Oppen}},\ }\href@noop {} {\bibfield  {journal} {\bibinfo  {journal} {Physical review letters}\ }\textbf {\bibinfo {volume} {105}},\ \bibinfo {pages} {177002} (\bibinfo {year} {2010})}\BibitemShut {NoStop}%
\bibitem [{\citenamefont {Lutchyn}\ \emph {et~al.}(2010)\citenamefont {Lutchyn}, \citenamefont {Sau},\ and\ \citenamefont {Das~Sarma}}]{lutchyn2010majorana-super-semi}%
  \BibitemOpen
  \bibfield  {author} {\bibinfo {author} {\bibfnamefont {R.~M.}\ \bibnamefont {Lutchyn}}, \bibinfo {author} {\bibfnamefont {J.~D.}\ \bibnamefont {Sau}},\ and\ \bibinfo {author} {\bibfnamefont {S.}~\bibnamefont {Das~Sarma}},\ }\href@noop {} {\bibfield  {journal} {\bibinfo  {journal} {Physical review letters}\ }\textbf {\bibinfo {volume} {105}},\ \bibinfo {pages} {077001} (\bibinfo {year} {2010})}\BibitemShut {NoStop}%
\bibitem [{\citenamefont {Pientka}\ \emph {et~al.}(2017{\natexlab{b}})\citenamefont {Pientka}, \citenamefont {Keselman}, \citenamefont {Berg}, \citenamefont {Yacoby}, \citenamefont {Stern},\ and\ \citenamefont {Halperin}}]{pientka2017mzm-planar-jj}%
  \BibitemOpen
  \bibfield  {author} {\bibinfo {author} {\bibfnamefont {F.}~\bibnamefont {Pientka}}, \bibinfo {author} {\bibfnamefont {A.}~\bibnamefont {Keselman}}, \bibinfo {author} {\bibfnamefont {E.}~\bibnamefont {Berg}}, \bibinfo {author} {\bibfnamefont {A.}~\bibnamefont {Yacoby}}, \bibinfo {author} {\bibfnamefont {A.}~\bibnamefont {Stern}},\ and\ \bibinfo {author} {\bibfnamefont {B.~I.}\ \bibnamefont {Halperin}},\ }\href@noop {} {\bibfield  {journal} {\bibinfo  {journal} {Physical Review X}\ }\textbf {\bibinfo {volume} {7}},\ \bibinfo {pages} {021032} (\bibinfo {year} {2017}{\natexlab{b}})}\BibitemShut {NoStop}%
\bibitem [{\citenamefont {Hell}\ \emph {et~al.}(2017)\citenamefont {Hell}, \citenamefont {Leijnse},\ and\ \citenamefont {Flensberg}}]{hell2017mzm-planar-jj}%
  \BibitemOpen
  \bibfield  {author} {\bibinfo {author} {\bibfnamefont {M.}~\bibnamefont {Hell}}, \bibinfo {author} {\bibfnamefont {M.}~\bibnamefont {Leijnse}},\ and\ \bibinfo {author} {\bibfnamefont {K.}~\bibnamefont {Flensberg}},\ }\href@noop {} {\bibfield  {journal} {\bibinfo  {journal} {Physical review letters}\ }\textbf {\bibinfo {volume} {118}},\ \bibinfo {pages} {107701} (\bibinfo {year} {2017})}\BibitemShut {NoStop}%
\bibitem [{\citenamefont {Ren}\ \emph {et~al.}(2019)\citenamefont {Ren}, \citenamefont {Pientka}, \citenamefont {Hart}, \citenamefont {Pierce}, \citenamefont {Kosowsky}, \citenamefont {Lunczer}, \citenamefont {Schlereth}, \citenamefont {Scharf}, \citenamefont {Hankiewicz}, \citenamefont {Molenkamp} \emph {et~al.}}]{ren2019topological-phase-bias-theory}%
  \BibitemOpen
  \bibfield  {author} {\bibinfo {author} {\bibfnamefont {H.}~\bibnamefont {Ren}}, \bibinfo {author} {\bibfnamefont {F.}~\bibnamefont {Pientka}}, \bibinfo {author} {\bibfnamefont {S.}~\bibnamefont {Hart}}, \bibinfo {author} {\bibfnamefont {A.~T.}\ \bibnamefont {Pierce}}, \bibinfo {author} {\bibfnamefont {M.}~\bibnamefont {Kosowsky}}, \bibinfo {author} {\bibfnamefont {L.}~\bibnamefont {Lunczer}}, \bibinfo {author} {\bibfnamefont {R.}~\bibnamefont {Schlereth}}, \bibinfo {author} {\bibfnamefont {B.}~\bibnamefont {Scharf}}, \bibinfo {author} {\bibfnamefont {E.~M.}\ \bibnamefont {Hankiewicz}}, \bibinfo {author} {\bibfnamefont {L.~W.}\ \bibnamefont {Molenkamp}}, \emph {et~al.},\ }\href@noop {} {\bibfield  {journal} {\bibinfo  {journal} {Nature}\ }\textbf {\bibinfo {volume} {569}},\ \bibinfo {pages} {93} (\bibinfo {year} {2019})}\BibitemShut {NoStop}%
\bibitem [{\citenamefont {Fornieri}\ \emph {et~al.}(2019)\citenamefont {Fornieri}, \citenamefont {Whiticar}, \citenamefont {Setiawan}, \citenamefont {Portol{\'e}s}, \citenamefont {Drachmann}, \citenamefont {Keselman}, \citenamefont {Gronin}, \citenamefont {Thomas}, \citenamefont {Wang}, \citenamefont {Kallaher} \emph {et~al.}}]{fornieri2019-topological-exp-min-re-entrant}%
  \BibitemOpen
  \bibfield  {author} {\bibinfo {author} {\bibfnamefont {A.}~\bibnamefont {Fornieri}}, \bibinfo {author} {\bibfnamefont {A.~M.}\ \bibnamefont {Whiticar}}, \bibinfo {author} {\bibfnamefont {F.}~\bibnamefont {Setiawan}}, \bibinfo {author} {\bibfnamefont {E.}~\bibnamefont {Portol{\'e}s}}, \bibinfo {author} {\bibfnamefont {A.~C.}\ \bibnamefont {Drachmann}}, \bibinfo {author} {\bibfnamefont {A.}~\bibnamefont {Keselman}}, \bibinfo {author} {\bibfnamefont {S.}~\bibnamefont {Gronin}}, \bibinfo {author} {\bibfnamefont {C.}~\bibnamefont {Thomas}}, \bibinfo {author} {\bibfnamefont {T.}~\bibnamefont {Wang}}, \bibinfo {author} {\bibfnamefont {R.}~\bibnamefont {Kallaher}}, \emph {et~al.},\ }\href@noop {} {\bibfield  {journal} {\bibinfo  {journal} {Nature}\ }\textbf {\bibinfo {volume} {569}},\ \bibinfo {pages} {89} (\bibinfo {year} {2019})}\BibitemShut {NoStop}%
\bibitem [{\citenamefont {Dartiailh}\ \emph {et~al.}(2021)\citenamefont {Dartiailh}, \citenamefont {Mayer}, \citenamefont {Yuan}, \citenamefont {Wickramasinghe}, \citenamefont {Matos-Abiague}, \citenamefont {{\v{Z}}uti{\'c}},\ and\ \citenamefont {Shabani}}]{dartiailh2021-min-Ic}%
  \BibitemOpen
  \bibfield  {author} {\bibinfo {author} {\bibfnamefont {M.~C.}\ \bibnamefont {Dartiailh}}, \bibinfo {author} {\bibfnamefont {W.}~\bibnamefont {Mayer}}, \bibinfo {author} {\bibfnamefont {J.}~\bibnamefont {Yuan}}, \bibinfo {author} {\bibfnamefont {K.~S.}\ \bibnamefont {Wickramasinghe}}, \bibinfo {author} {\bibfnamefont {A.}~\bibnamefont {Matos-Abiague}}, \bibinfo {author} {\bibfnamefont {I.}~\bibnamefont {{\v{Z}}uti{\'c}}},\ and\ \bibinfo {author} {\bibfnamefont {J.}~\bibnamefont {Shabani}},\ }\href@noop {} {\bibfield  {journal} {\bibinfo  {journal} {Physical Review Letters}\ }\textbf {\bibinfo {volume} {126}},\ \bibinfo {pages} {036802} (\bibinfo {year} {2021})}\BibitemShut {NoStop}%
\end{thebibliography}%


\begin{thebibliography}{4}%
\makeatletter
\providecommand \@ifxundefined [1]{%
 \@ifx{#1\undefined}
}%
\providecommand \@ifnum [1]{%
 \ifnum #1\expandafter \@firstoftwo
 \else \expandafter \@secondoftwo
 \fi
}%
\providecommand \@ifx [1]{%
 \ifx #1\expandafter \@firstoftwo
 \else \expandafter \@secondoftwo
 \fi
}%
\providecommand \natexlab [1]{#1}%
\providecommand \enquote  [1]{``#1''}%
\providecommand \bibnamefont  [1]{#1}%
\providecommand \bibfnamefont [1]{#1}%
\providecommand \citenamefont [1]{#1}%
\providecommand \href@noop [0]{\@secondoftwo}%
\providecommand \href [0]{\begingroup \@sanitize@url \@href}%
\providecommand \@href[1]{\@@startlink{#1}\@@href}%
\providecommand \@@href[1]{\endgroup#1\@@endlink}%
\providecommand \@sanitize@url [0]{\catcode `\\12\catcode `\$12\catcode `\&12\catcode `\#12\catcode `\^12\catcode `\_12\catcode `\%12\relax}%
\providecommand \@@startlink[1]{}%
\providecommand \@@endlink[0]{}%
\providecommand \url  [0]{\begingroup\@sanitize@url \@url }%
\providecommand \@url [1]{\endgroup\@href {#1}{\urlprefix }}%
\providecommand \urlprefix  [0]{URL }%
\providecommand \Eprint [0]{\href }%
\providecommand \doibase [0]{https://doi.org/}%
\providecommand \selectlanguage [0]{\@gobble}%
\providecommand \bibinfo  [0]{\@secondoftwo}%
\providecommand \bibfield  [0]{\@secondoftwo}%
\providecommand \translation [1]{[#1]}%
\providecommand \BibitemOpen [0]{}%
\providecommand \bibitemStop [0]{}%
\providecommand \bibitemNoStop [0]{.\EOS\space}%
\providecommand \EOS [0]{\spacefactor3000\relax}%
\providecommand \BibitemShut  [1]{\csname bibitem#1\endcsname}%
\let\auto@bib@innerbib\@empty
\bibitem [{\citenamefont {Shabani}\ \emph {et~al.}(2016)\citenamefont {Shabani}, \citenamefont {Kj{\ae}rgaard}, \citenamefont {Suominen}, \citenamefont {Kim}, \citenamefont {Nichele}, \citenamefont {Pakrouski}, \citenamefont {Stankevic}, \citenamefont {Lutchyn}, \citenamefont {Krogstrup}, \citenamefont {Feidenhans'l} \emph {et~al.}}]{shabani2016AlInAs}%
  \BibitemOpen
  \bibfield  {author} {\bibinfo {author} {\bibfnamefont {J.}~\bibnamefont {Shabani}}, \bibinfo {author} {\bibfnamefont {M.}~\bibnamefont {Kj{\ae}rgaard}}, \bibinfo {author} {\bibfnamefont {H.~J.}\ \bibnamefont {Suominen}}, \bibinfo {author} {\bibfnamefont {Y.}~\bibnamefont {Kim}}, \bibinfo {author} {\bibfnamefont {F.}~\bibnamefont {Nichele}}, \bibinfo {author} {\bibfnamefont {K.}~\bibnamefont {Pakrouski}}, \bibinfo {author} {\bibfnamefont {T.}~\bibnamefont {Stankevic}}, \bibinfo {author} {\bibfnamefont {R.~M.}\ \bibnamefont {Lutchyn}}, \bibinfo {author} {\bibfnamefont {P.}~\bibnamefont {Krogstrup}}, \bibinfo {author} {\bibfnamefont {R.}~\bibnamefont {Feidenhans'l}}, \emph {et~al.},\ }\href@noop {} {\bibfield  {journal} {\bibinfo  {journal} {Physical Review B}\ }\textbf {\bibinfo {volume} {93}},\ \bibinfo {pages} {155402} (\bibinfo {year} {2016})}\BibitemShut {NoStop}%
\bibitem [{\citenamefont {Wickramasinghe}\ \emph {et~al.}(2018)\citenamefont {Wickramasinghe}, \citenamefont {Mayer}, \citenamefont {Yuan}, \citenamefont {Nguyen}, \citenamefont {Jiao}, \citenamefont {Manucharyan},\ and\ \citenamefont {Shabani}}]{wickramasinghe2018transport-AlInAs}%
  \BibitemOpen
  \bibfield  {author} {\bibinfo {author} {\bibfnamefont {K.~S.}\ \bibnamefont {Wickramasinghe}}, \bibinfo {author} {\bibfnamefont {W.}~\bibnamefont {Mayer}}, \bibinfo {author} {\bibfnamefont {J.}~\bibnamefont {Yuan}}, \bibinfo {author} {\bibfnamefont {T.}~\bibnamefont {Nguyen}}, \bibinfo {author} {\bibfnamefont {L.}~\bibnamefont {Jiao}}, \bibinfo {author} {\bibfnamefont {V.}~\bibnamefont {Manucharyan}},\ and\ \bibinfo {author} {\bibfnamefont {J.}~\bibnamefont {Shabani}},\ }\href@noop {} {\bibfield  {journal} {\bibinfo  {journal} {Applied Physics Letters}\ }\textbf {\bibinfo {volume} {113}} (\bibinfo {year} {2018})}\BibitemShut {NoStop}%
\bibitem [{\citenamefont {Strickland}\ \emph {et~al.}(2022)\citenamefont {Strickland}, \citenamefont {Hatefipour}, \citenamefont {Langone}, \citenamefont {Farzaneh},\ and\ \citenamefont {Shabani}}]{strickland2022controlling}%
  \BibitemOpen
  \bibfield  {author} {\bibinfo {author} {\bibfnamefont {W.~M.}\ \bibnamefont {Strickland}}, \bibinfo {author} {\bibfnamefont {M.}~\bibnamefont {Hatefipour}}, \bibinfo {author} {\bibfnamefont {D.}~\bibnamefont {Langone}}, \bibinfo {author} {\bibfnamefont {S.}~\bibnamefont {Farzaneh}},\ and\ \bibinfo {author} {\bibfnamefont {J.}~\bibnamefont {Shabani}},\ }\href@noop {} {\bibfield  {journal} {\bibinfo  {journal} {Applied Physics Letters}\ }\textbf {\bibinfo {volume} {121}} (\bibinfo {year} {2022})}\BibitemShut {NoStop}%
\bibitem [{\citenamefont {Lankhorst}\ \emph {et~al.}(2018)\citenamefont {Lankhorst}, \citenamefont {Brinkman}, \citenamefont {Hilgenkamp}, \citenamefont {Poccia},\ and\ \citenamefont {Golubov}}]{lankhorst2018annealed}%
  \BibitemOpen
  \bibfield  {author} {\bibinfo {author} {\bibfnamefont {M.}~\bibnamefont {Lankhorst}}, \bibinfo {author} {\bibfnamefont {A.}~\bibnamefont {Brinkman}}, \bibinfo {author} {\bibfnamefont {H.}~\bibnamefont {Hilgenkamp}}, \bibinfo {author} {\bibfnamefont {N.}~\bibnamefont {Poccia}},\ and\ \bibinfo {author} {\bibfnamefont {A.}~\bibnamefont {Golubov}},\ }\href@noop {} {\bibfield  {journal} {\bibinfo  {journal} {Condensed Matter}\ }\textbf {\bibinfo {volume} {3}},\ \bibinfo {pages} {19} (\bibinfo {year} {2018})}\BibitemShut {NoStop}%
\end{thebibliography}%
\end{document}


\title{Supplementary Information: }

\author{Melissa Mikalsen}
\email{mem9982@nyu.edu}
\affiliation{Center for Quantum Information Physics, Department of Physics, New York University, New York, NY 10003, USA}

\author{Alexander-Georg Penner}
\affiliation{Dahlem Center for Complex Quantum Systems and Fachbereich Physik, Freie Universität Berlin, 14195 Berlin, Germany}

\author{Samuel D. Escribano}
\affiliation{Department of Condensed Matter Physics, Weizmann Institute of Science, Rehovot 7610001, Israel}

\author{Nadav Drechsler}
\affiliation{Department of Condensed Matter Physics, Weizmann Institute of Science, Rehovot 7610001, Israel}

\author{Arunav Bordoloi}
\affiliation{Center for Quantum Information Physics, Department of Physics, New York University, New York, NY 10003, USA}

\author{Jacob Issokson}
\affiliation{Center for Quantum Information Physics, Department of Physics, New York University, New York, NY 10003, USA}

\author{Yuval Oreg}
\affiliation{Department of Condensed Matter Physics, Weizmann Institute of Science, Rehovot 7610001, Israel}

\author{Javad Shabani}
\email{corresponding author, jshabani@nyu.edu}
\affiliation{Center for Quantum Information Physics, Department of Physics, New York University, New York, NY 10003, USA}

\date{\today}
\maketitle

\section{\label{sec:material}Material heterostructure}
The heterostructure consists of an InAs 2-dimensional electron gas (2DEG) grown by molecular beam epitaxy capped in-situ with an aluminum layer. Growth was carried out on an epi-ready, semi-insulating (Fe-doped), 350 $\mu$m thick, 2 inch-diameter, single-side polished InP wafer (JX Nippon Mining \& Metals Corporation-Acrotec). Before growth, the native oxide was thermally desorbed under an arsenic overpressure in an ultrahigh vacuum chamber. A superlattice of alternating layers of In$_{0.52}$Al$_{0.48}$As and In$_{0.53}$Ga$_{0.47}$As was first grown. This was followed by a graded buffer layer designed to minimize the compressive strain on the active region. The buffer layer consists of a 400 nm thick  In$_{x}$Al$_{1-x}$As layer, where the composition of $x$ was graded in steps of $\Delta x = 0.02$ every 25 nm. The quantum well was then formed by a 4 nm bottom In$_{0.81}$Ga$_{0.19}$As barrier, a 4 nm InAs layer, and a 10 nm In$_{0.81}$Ga$_{0.19}$As top barrier. The structure is delta-doped 6 nm below the active region with Si. The wafer was then cooled below 0${^\circ}$C for the in-situ deposition of a 20 nm thick Al layer. Further information on the growth procedure can be found in Refs.~\citenum{shabani2016AlInAs,wickramasinghe2018transport-AlInAs,strickland2022controlling}.

In low temperature magnetotransport measurements of a van der Pauw sample of a different piece of the same wafer we fabricated the devices on, we measured the average electron mobitly to be $\mu = 1.56 \times 10^4$ cm$^2$/Vs and the density to be $n = 1.11\times 10^{12}$ cm$^{-2}$, leading to the mean free path of $l_e = 271$ nm. 


\section{\label{sec:fab}Device nanofabrication}

Both the hexagonal and the square junction arrays were fabricated on two 5 mm x 5 mm pieces of the same wafer. The junction array devices were patterned with electron beam lithography and defined by selectively wet etching the Al with Transene Type D Al etchant. The mesa features were subsequently patterned with photo-lithography and defined by wet-etching the Al with Transene Type D, followed by wet-etching the InAlAs and InGaAs layers with a 1:1:40 volumetric ratio of phosphoric acid (85\%), hydrogen peroxide (30\%), and deionized water.

\section{Numerical details}

We model the Josephson junction array in the experiment through a frustrated 2d xy model as follows, 
\begin{equation}
    H = -E_J\sum_{\langle ij\rangle} \cos(\varphi_i -\varphi_j -2\pi aA_{ij}/\Phi_0).
    \label{eq: xy model}
\end{equation}

We search for ground-state configurations of the array as described by Eq. \eqref{eq: xy model}. The dynamics is simulated using the RCSJ model following the algorithm described in Ref. \cite{lankhorst2018annealed}. The Kirchhoff rules at each lattice site result in $6N^2$ coupled linear differential equations, with $N \times N$ being the number of unit cells. Nonzero temperature enters the dynamics via Langevin currents (Johnson-Nyquist noise). The algorithm evolves the state of the system towards a minimum-energy configuration by gradually decreasing the temperature. 

The magnetic-flux configuration is encoded in the vector potentials $A_{ij}$ in Eq. \eqref{eq: xy model}.  Unless specified otherwise, the parameters used in the simulations are: $\tau = 0.2/E_\text{J}$, $T= 0.5E_\text{J}$, the time discretization is written in dimensionless quantities $T_0=2\pi k_{\text{BT}}/E_{\text{J}}$, $\tau= (2eI_{\text{c}}R_{\text{n}}/ \hbar)t$\cite{lankhorst2018annealed}.

\begin{figure}[H]
    \centering
    \includegraphics[width=0.75\linewidth]{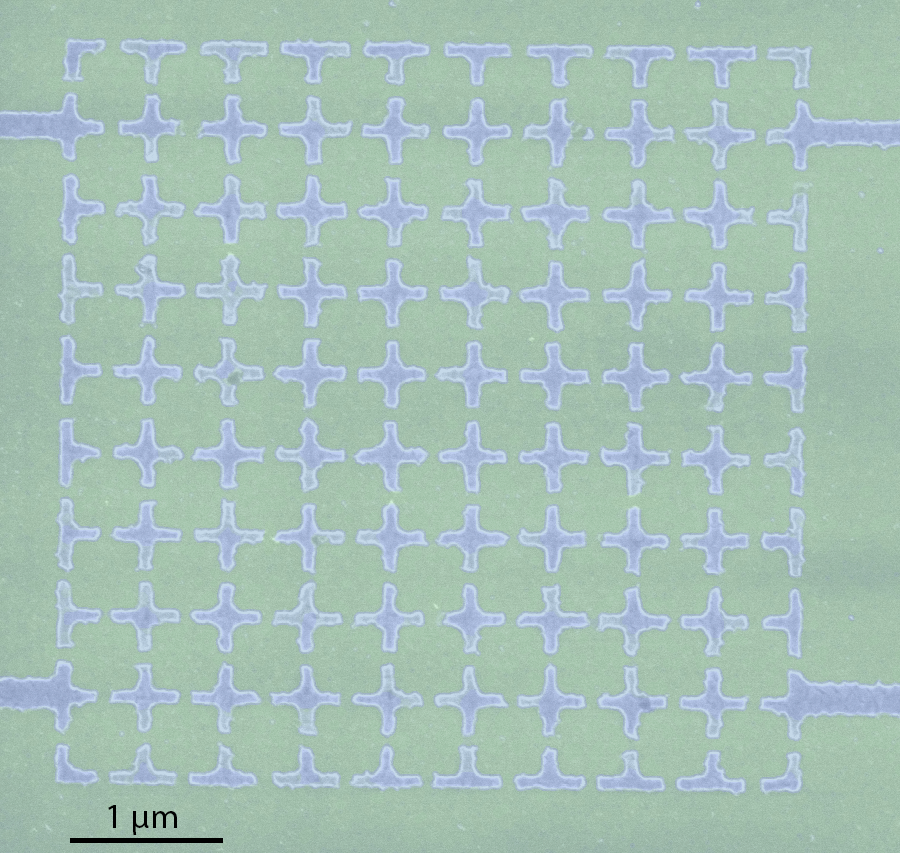}
    \caption{\textbf{Full square array.} Scanning electron micrograph of a of a representative device for the square array. The superconductor (blue) and semiconductor (green) are falsely colored and the scale bar is 1 $\mu$m. Bias current is injected from top left contact, grounded on top right contact and voltage is measured from bottom left to bottom right contacts with a lock-in.}
    \label{fig:full-square-array}
\end{figure}

\begin{figure}[H]
    \centering
    \includegraphics[width=0.75\linewidth]{supplement-figs/4362_hex-t=t'-zoom-2.png}
    \caption{\textbf{Full super-honeycomb array.} Scanning electron micrograph of a of a representative device for the super-honeycomb array. The superconductor (blue) and semiconductor (green) are falsely colored and the scale bar is 1 $\mu$m.  Bias current is injected from top left contact, grounded on top right contact and voltage is measured from bottom left to bottom right contacts with a lock-in. }
    \label{fig:full-hex-array}
\end{figure}

\begin{figure}[H]
    \centering
    \includegraphics[width=0.75\linewidth]{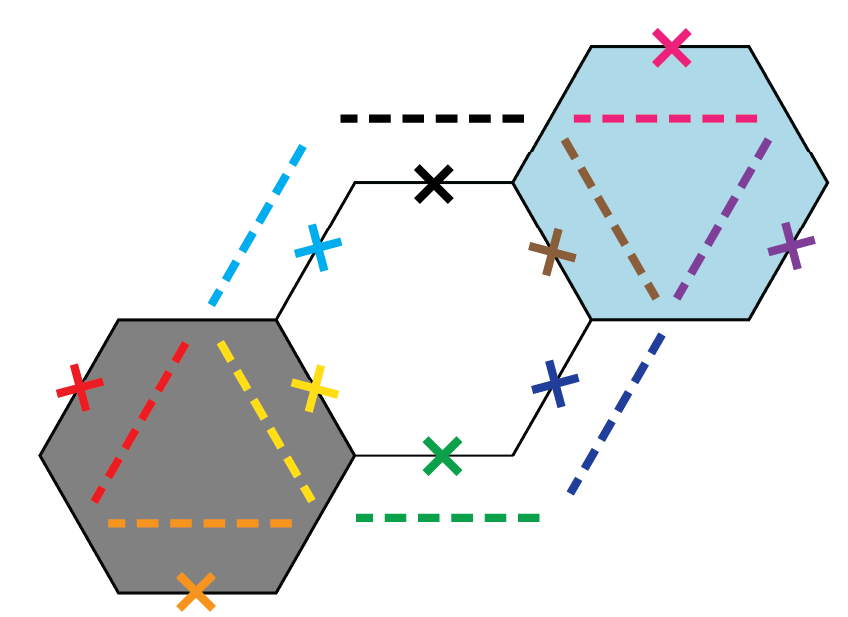}
    \caption{\textbf{Translation of junction lattice to kagome lattice.} The JJs (represented as crosses) are color-matched to and re-formed as edges of a kagome lattice. }
    \label{fig:kagome}
\end{figure}

\begin{figure}[H]
    \centering
    \includegraphics{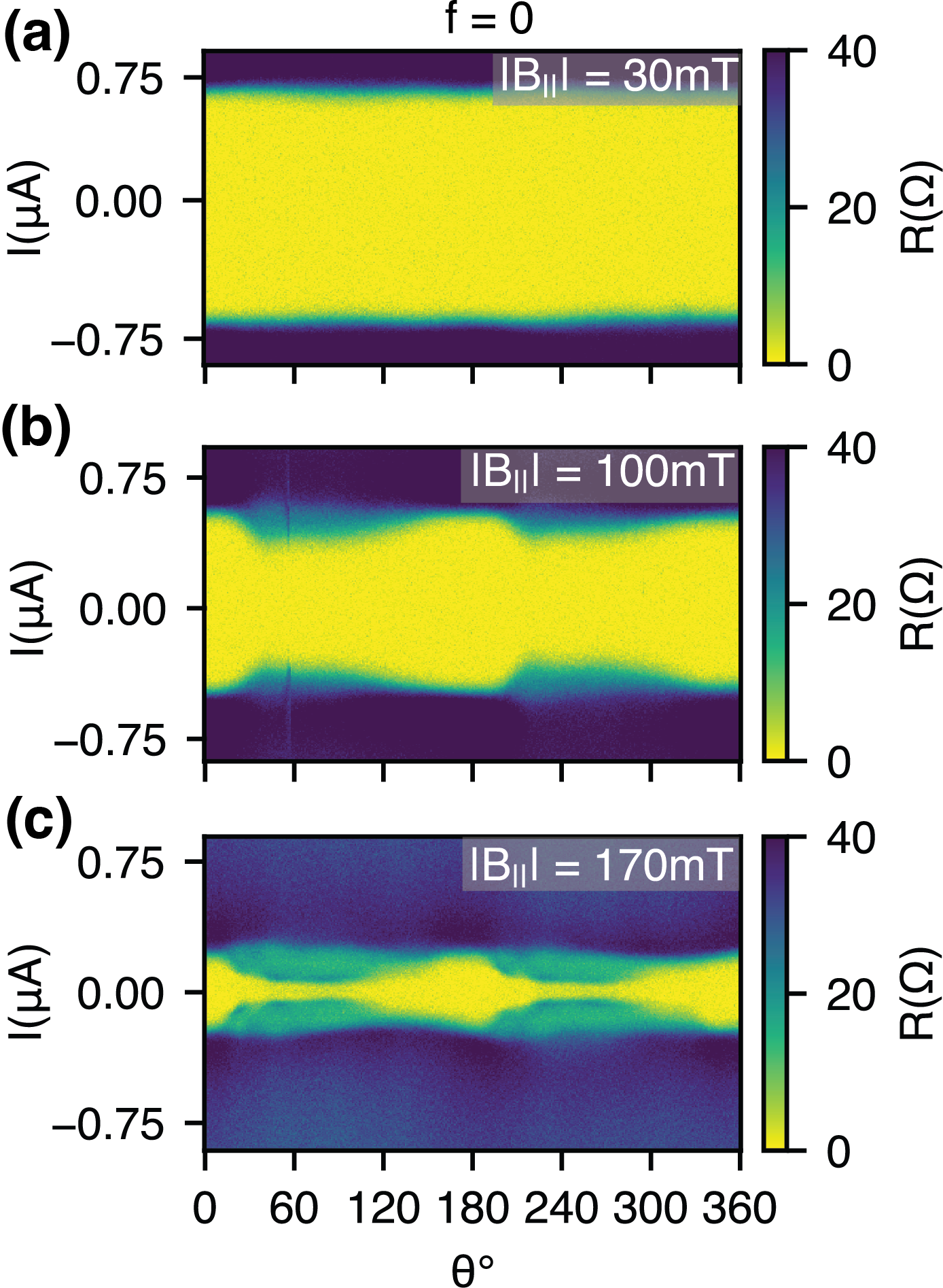}
    \caption{\textbf{Parallel magnetic field angle dependence of critical current at f =0 for hexagonal array. }Resistance as a function of current bias, I($\mu$A), and angle of parallel magnetic field, $\theta$, for array B, with in-plane magnitudes of 30 mT \textbf{(a)}, 100 mT \textbf{(b)}, and 170 mT\textbf{(c)}}
    \label{fig:f-0-angle}
\end{figure}

\begin{figure}[H]
    \centering
    \includegraphics{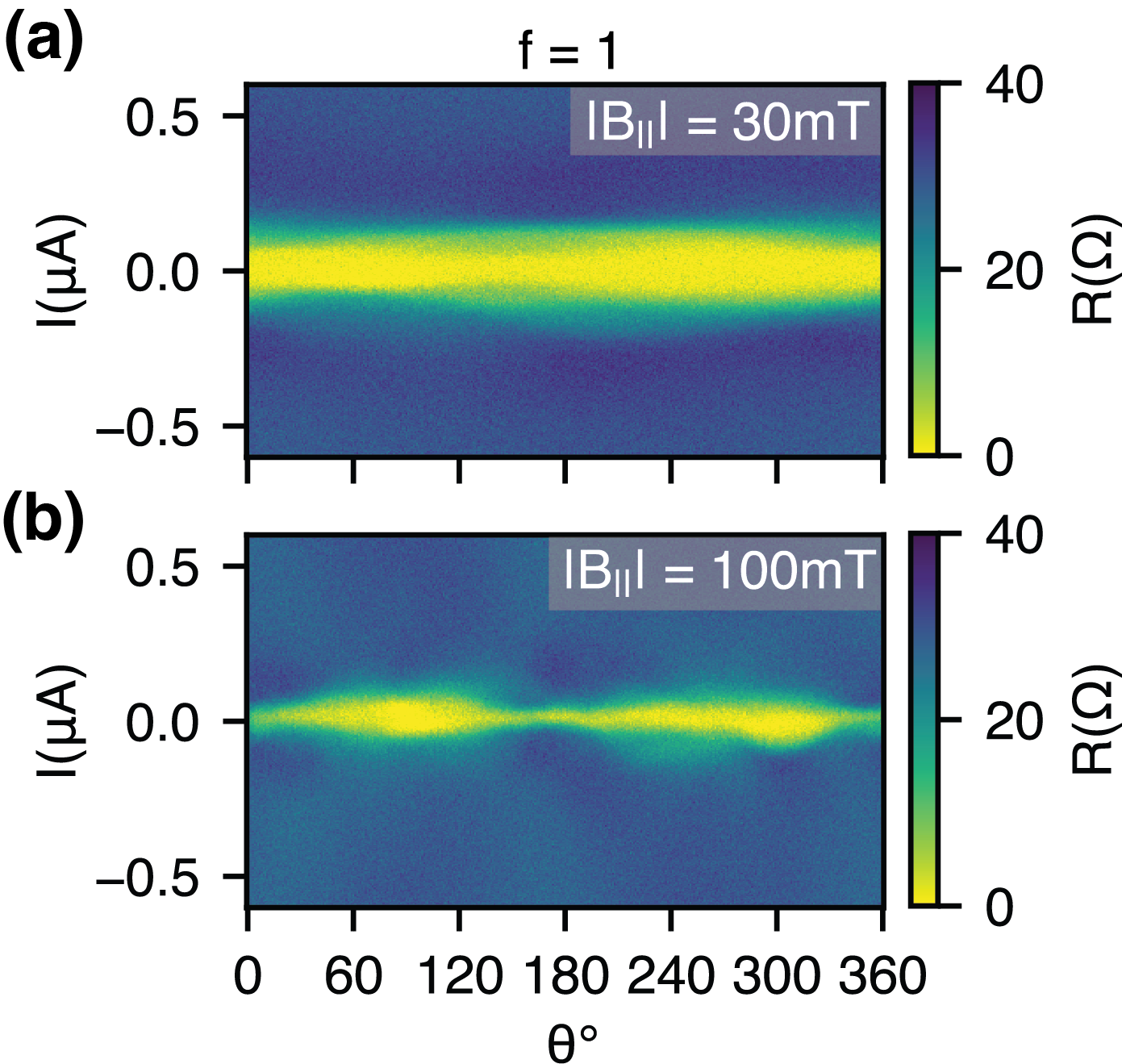}
    \caption{\textbf{Parallel magnetic field angle dependence of critical current at f =1. } Resistance as a function of current bias, I($\mu$A), and angle of parallel magnetic field, $\theta$, for array B, with in-plane magnitudes of 30 mT \textbf{(a)}, and 100 mT \textbf{(b)}. }
    \label{fig:f-1-angle}
\end{figure}

\begin{figure}[H]
    \centering
    \includegraphics{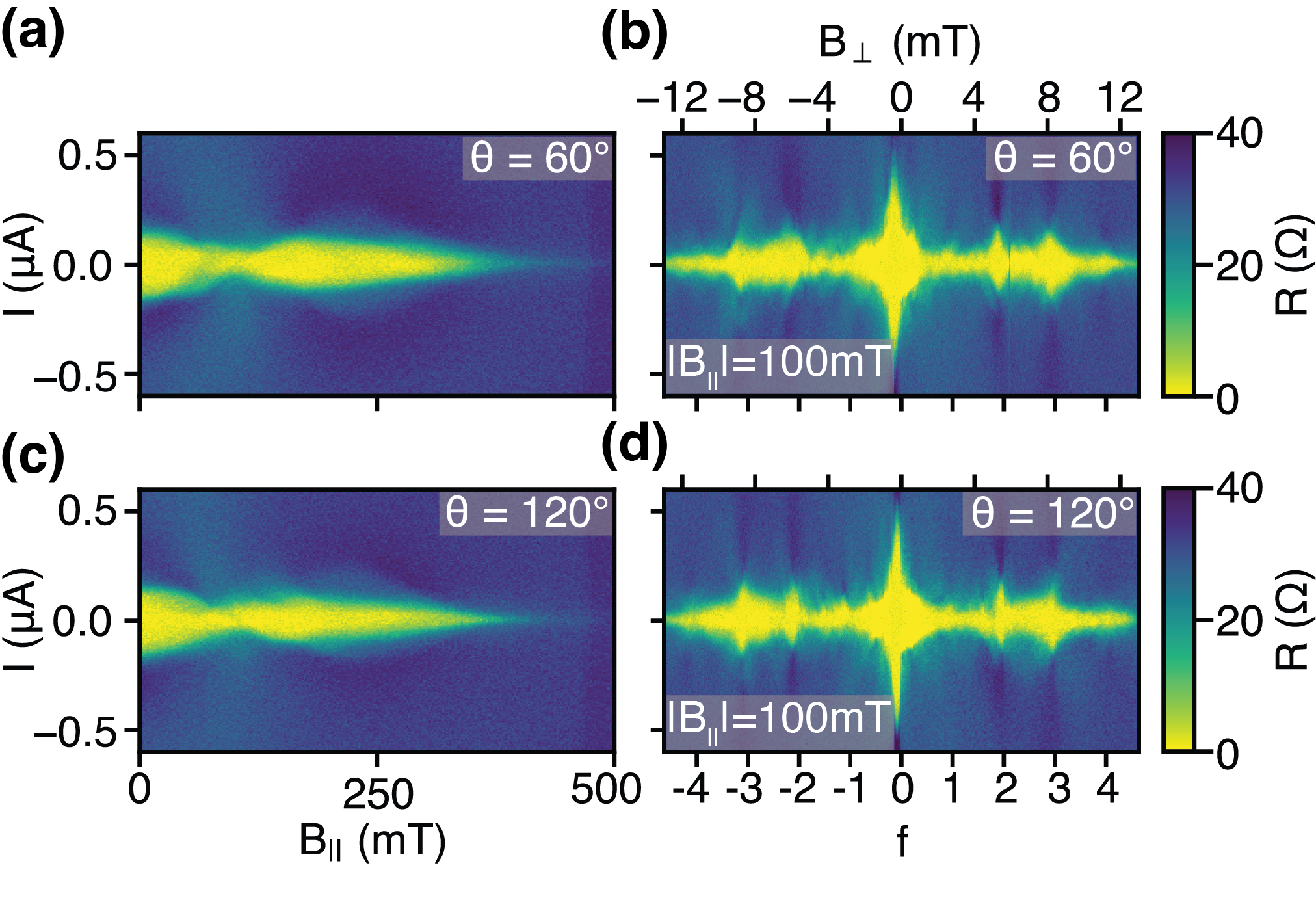}
    \caption{\textbf{Hexagonal Josephson junction arrays in parallel magnetic field at f =1.}Resistance as a function of current bias (I ($\mu$A)) and parallel magnetic field pointing along $\theta = 60^\circ$ \textbf{(a)} and $\theta = 120^\circ$ \textbf{(c)} at constant perpendicular magnetic field corresponding to f  = 1. Resistance as a function of current bias (I ($\mu$A)) and perpendicular magnetic field (B$_\perp$ (mT)) at constant parallel magnetic field, |B$_{||}$|= 100mT.}
    \label{fig:f-1-60and120deg}
\end{figure}

\begin{figure}[H]
    \centering
    \includegraphics{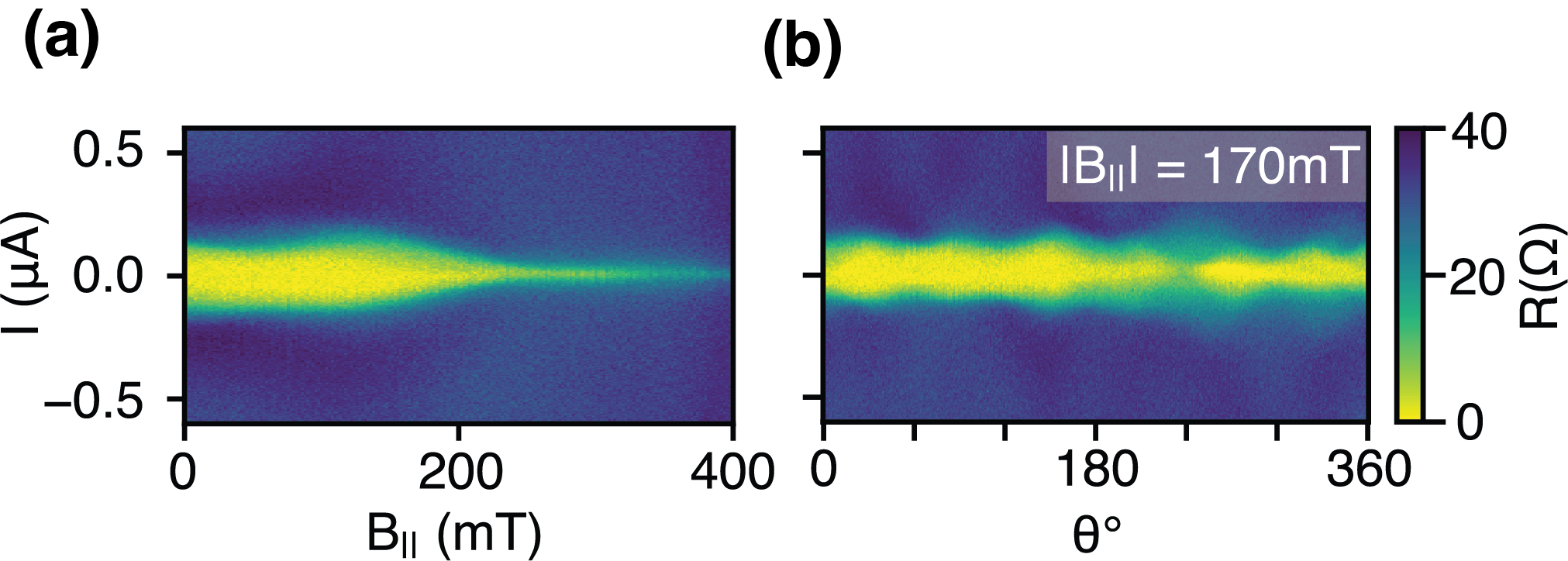}
    \caption{\textbf{Hexagonal array critical current dependence on parallel magnetic field magnitude and angle at f =2. } Resistance as a function of current bias (I ($\mu$A)) and parallel magnetic field pointing along $\theta = 0^\circ$ at constant perpendicular magnetic field corresponding to f  = 2 \textbf{(a)}. Resistance as a function of current bias, I($\mu$A), and angle of parallel magnetic field, $\theta^\circ $ with in-plane magnitude of 170 mT \textbf{(b)}, and constant perpendicular magnetic field corresponding to f =2. }
    \label{fig:f-2-angle}
\end{figure}

\begin{figure}[H]
    \centering
    \includegraphics{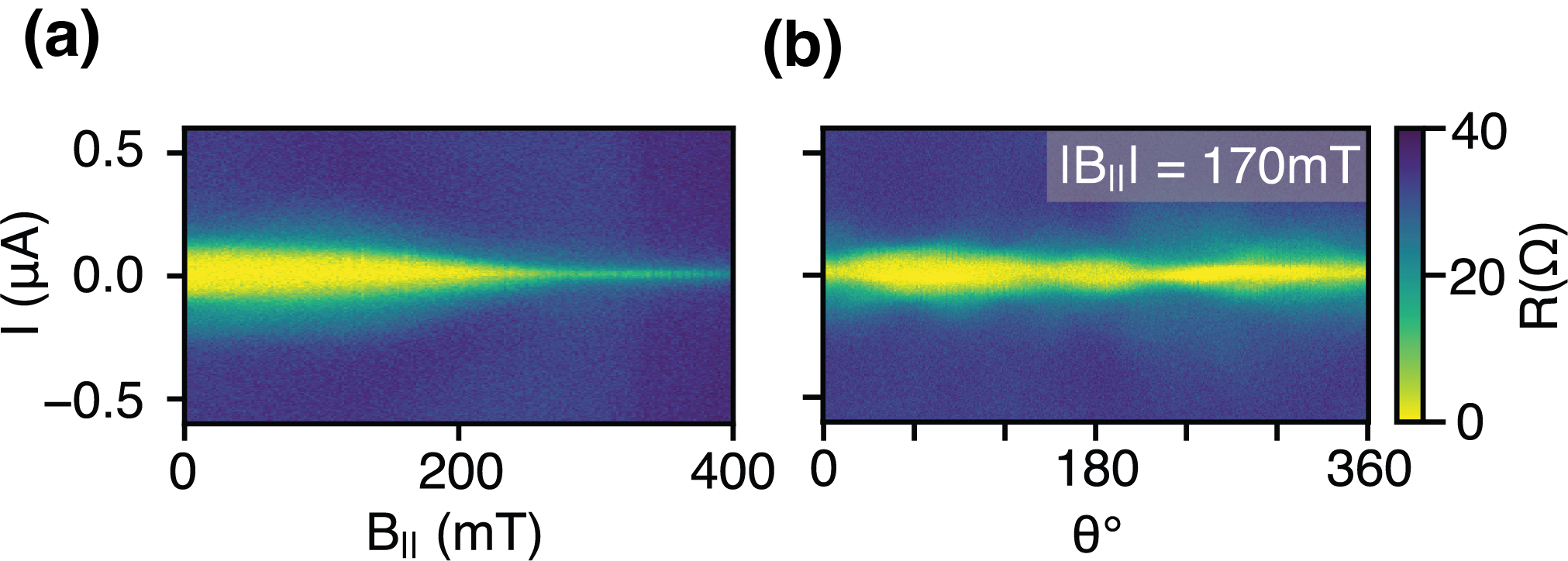}
    \caption{\textbf{Hexagonal array critical current dependence on parallel magnetic field magnitude and angle at f =3. }Resistance as a function of current bias (I ($\mu$A)) and parallel magnetic field pointing along $\theta = 0^\circ$ at constant perpendicular magnetic field corresponding to f  = 3 \textbf{(a)}. Resistance as a function of current bias, I($\mu$A), and angle of parallel magnetic field, $\theta^\circ$ with in-plane magnitude of 170 mT, and constant perpendicular magnetic field corresponding to f =3 \textbf{(b)}. }
    \label{fig:f-3-angle}
\end{figure}

\begin{figure}[H]
    \centering
    \includegraphics{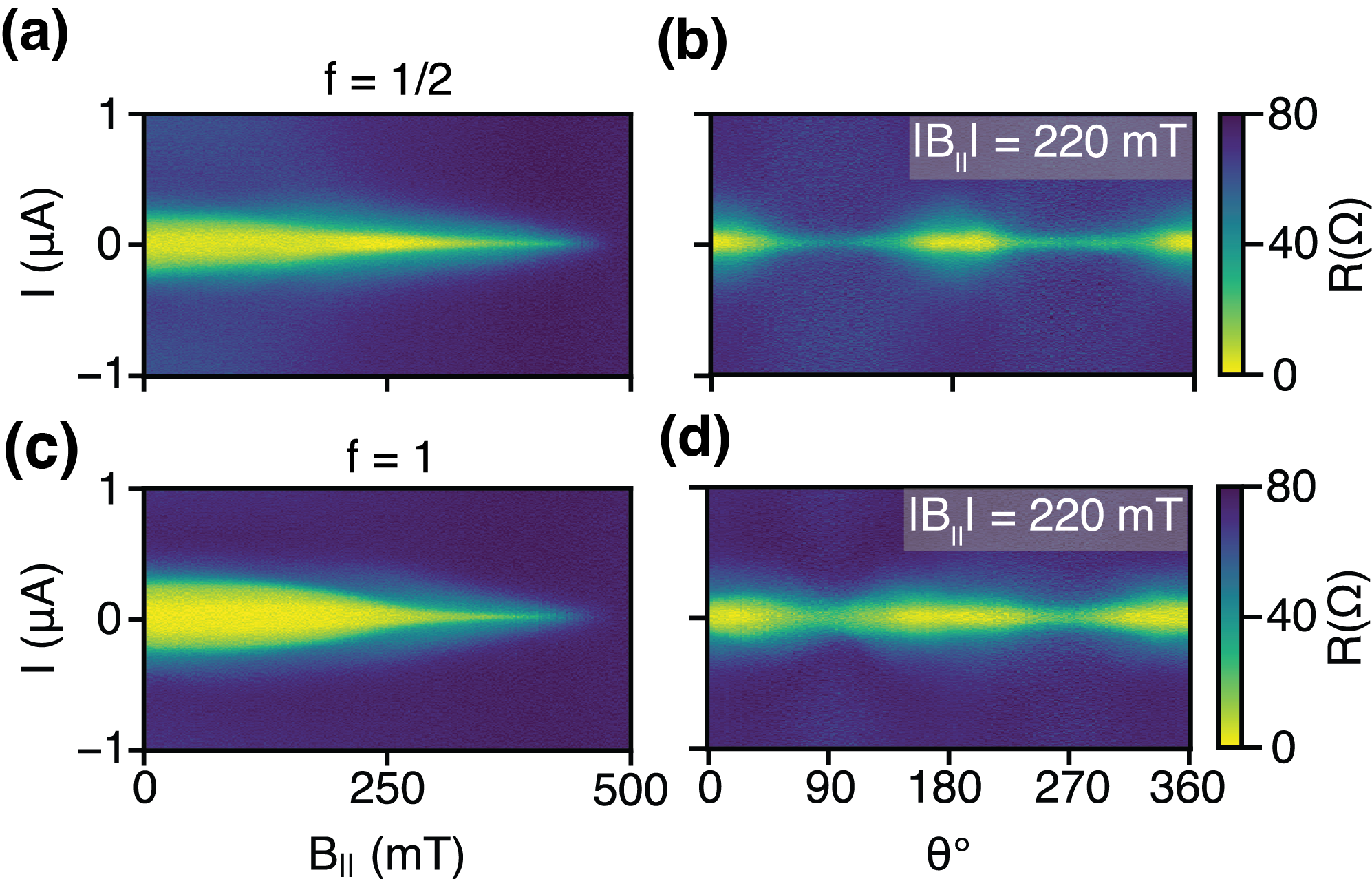}
    \caption{\textbf{Square array critical current dependence on parallel magnetic field magnitude and angle. }Resistance as a function of current bias (I ($\mu$A)) and parallel magnetic field pointing along $\theta = 0^\circ$ at constant perpendicular magnetic field corresponding to f  = 1/2 \textbf{(a)} and f  = 1 \textbf{(c)}. Resistance as a function of current bias, I($\mu$A), and angle of parallel magnetic field, $\theta^\circ$ with in-plane magnitudes of 220 mT, and constant perpendicular magnetic field corresponding to f =1/2 \textbf{(a)}, and f = 1 \textbf{(d)}. }
    \label{fig:square-angle}
\end{figure}

\begin{figure}[H]
    \centering
    \includegraphics{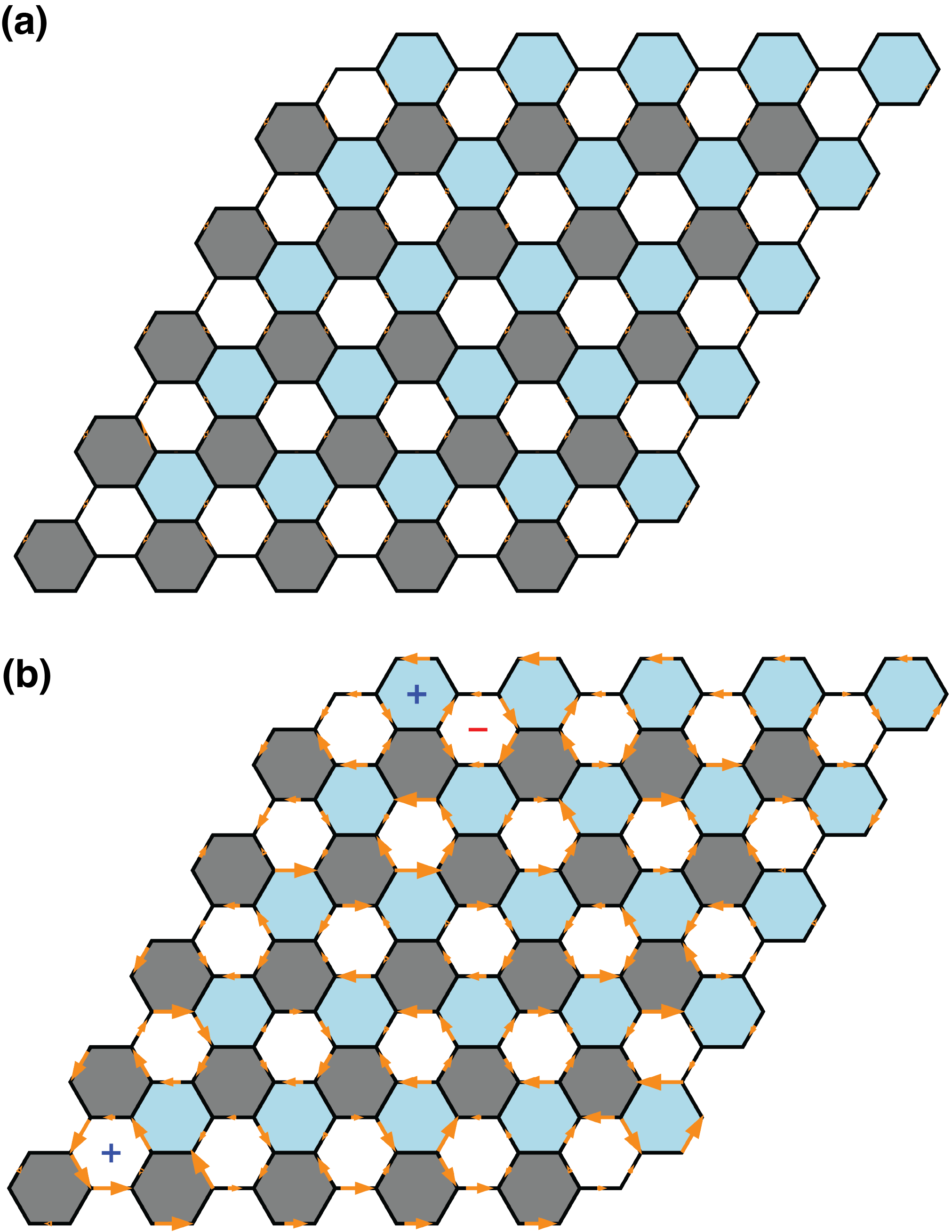}
    \caption{\textbf{Super-honeycomb array current numerical calculations f = 0. }Numerical calculations for f = 0 at $B_{||} = 0$ \textbf{(a)} and $B_{||} = 0.6$ \textbf{(b)}.}
    \label{fig:theory-f-0-current}
\end{figure}

\begin{figure}[H]
    \centering
    \includegraphics{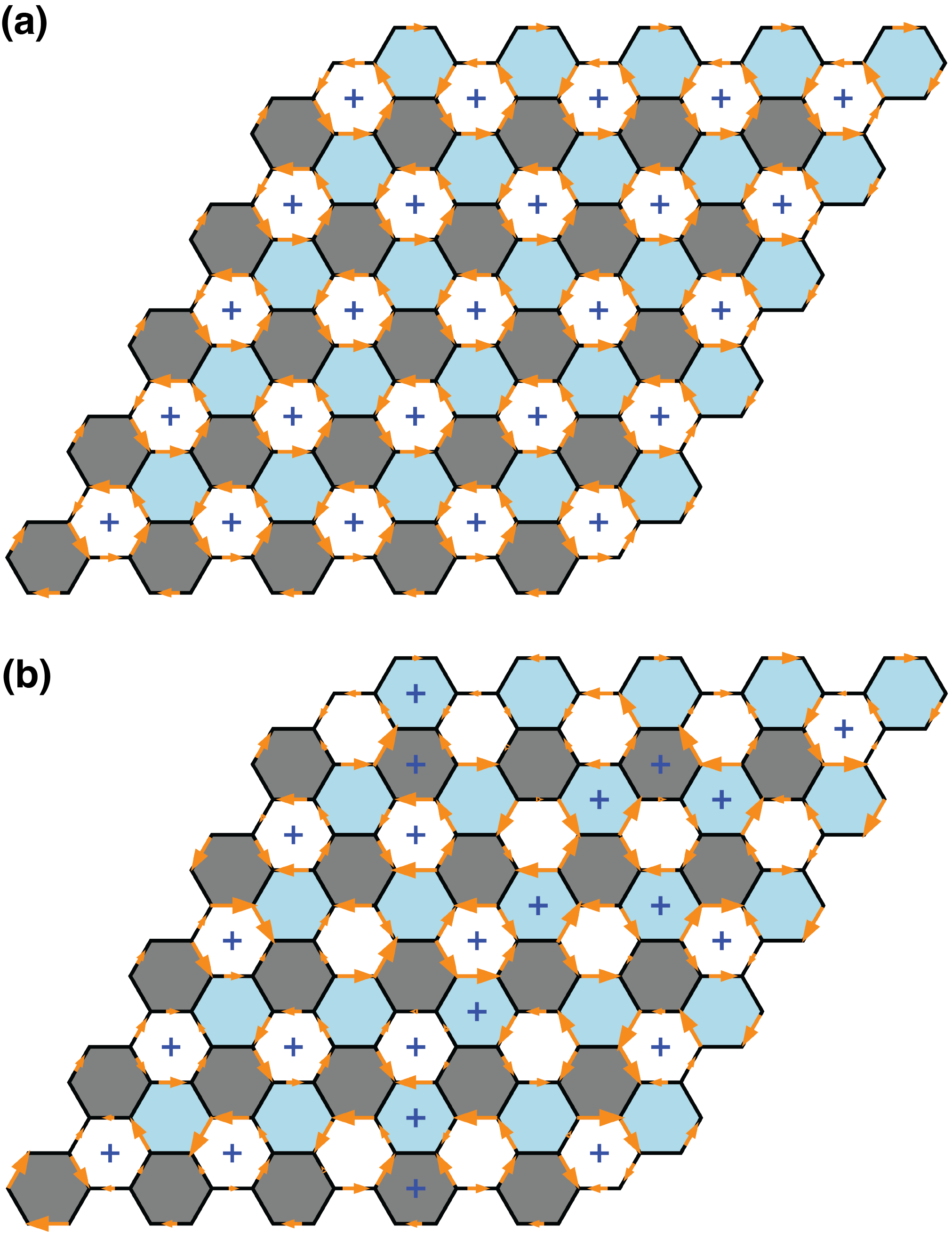}
    \caption{\textbf{Super-honeycomb array current numerical calculations f = 1. }Numerical calculations for f = 1 at $B_{||} = 0$ \textbf{(a)} and $B_{||} = 0.6$ \textbf{(b)}.}
    \label{fig:theory-f-1-current}
\end{figure}

\begin{figure}[H]
    \centering
    \includegraphics{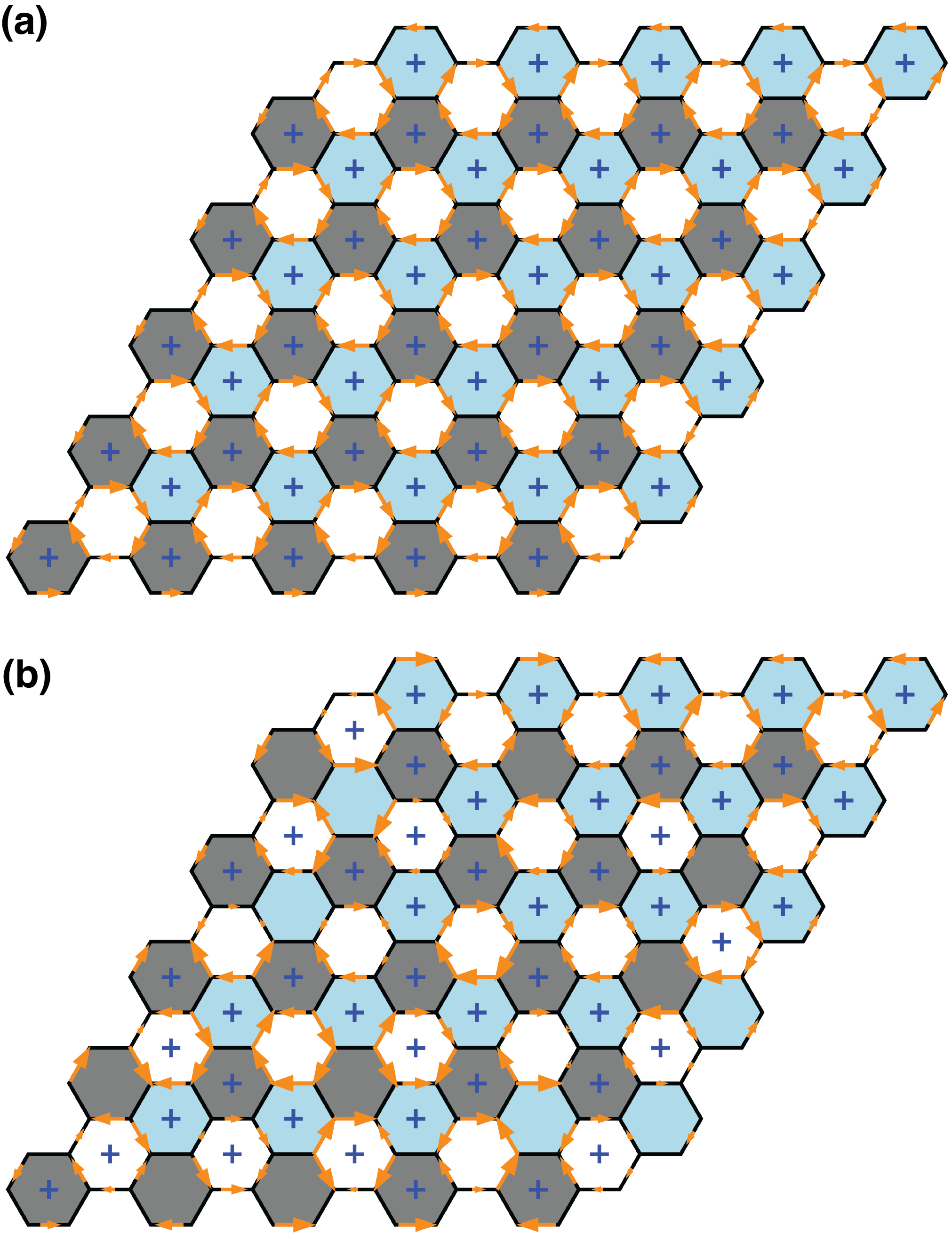}
    \caption{\textbf{Super-honeycomb array current numerical calculations f = 2. }Numerical calculations for f = 2 at $B_{||} = 0$ \textbf{(a)} and $B_{||} = 0.6$ \textbf{(b)}. }
    \label{fig:theory-f-2-current}
\end{figure}

\begin{figure}[H]
    \centering
    \includegraphics[width=1\linewidth]{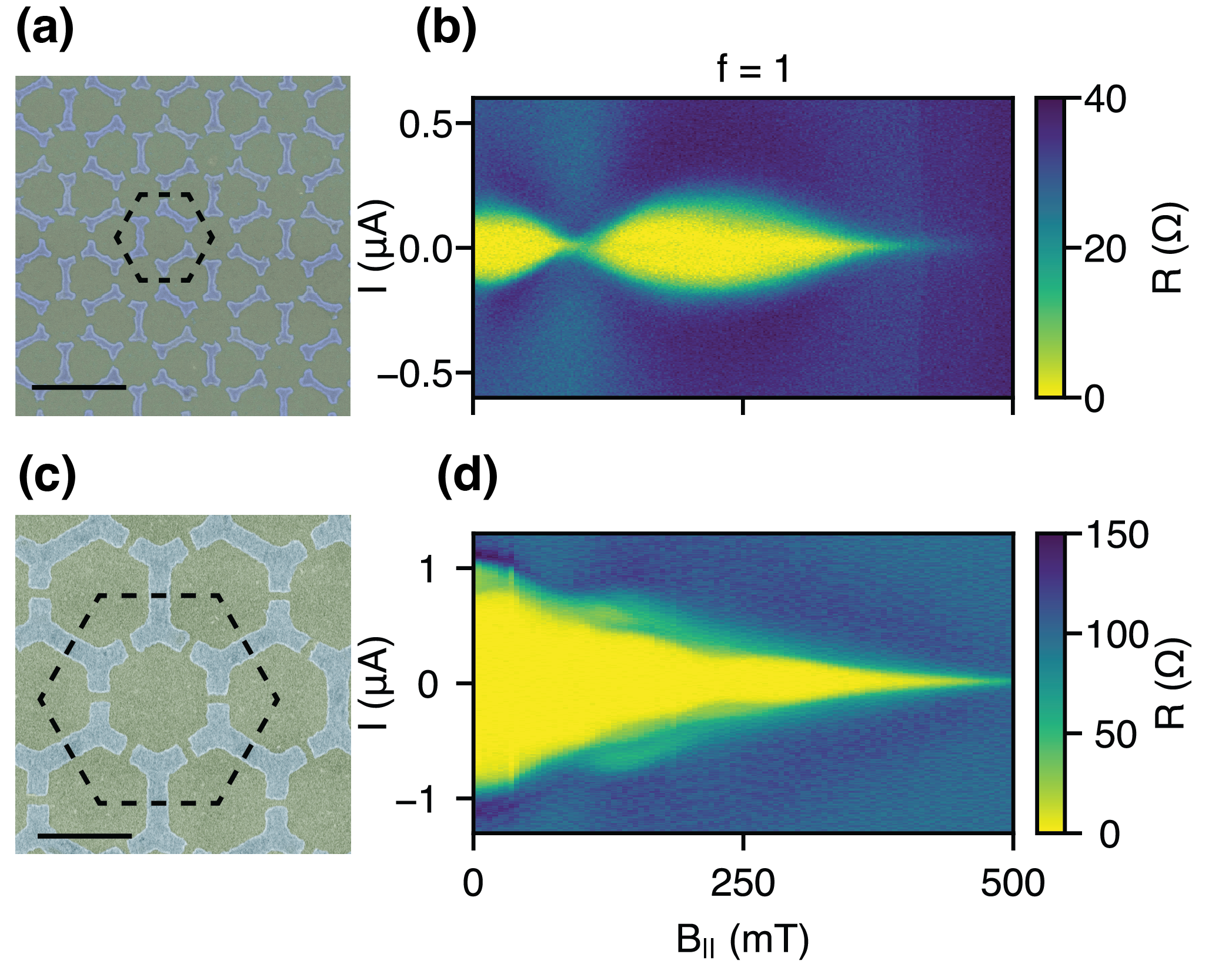}
    \caption{\textbf{Super-honeycomb array with lager unit cell area.} Scanning electron micrograph of a of a representative device for the super-honeycomb array with area (outlined in dashed black line) estimated to be 0.738 $\mu m^2$ \textbf{(a)} and 4.39 $\mu m^2$ \textbf{(c)} from SEM images. The superconductor (blue) and semiconductor (green) are falsely colored and the scale bar is 1 $\mu m$. Resistance as a function of current bias (I ($\mu$A)) and parallel magnetic field pointing along $\theta = 0^\circ$  for $A_{\text{UC}} = 0.738$ $\mu m^2$ \textbf{(b)} and $A_{\text{UC}} = 4.39$ $\mu m^2$\textbf{(d)} at constant perpendicular magnetic field corresponding to f  = 1. }
    \label{fig:vary-uc}
\end{figure}

\bibliographystyle{apsrev4-2}
\bibliography{bibs/bib}